\documentclass[%
 reprint,
superscriptaddress,
 amsmath,amssymb,
 aps,
]{revtex4-2}

\usepackage{graphicx}
\usepackage{dcolumn}
\usepackage{bm,}
\usepackage{dcolumn, comment}
\usepackage{array,multirow}
\usepackage{hyperref}
\usepackage{xcolor}
\usepackage{braket}
\usepackage{subfiles}

\begin{document}

\preprint{APS/123-QED}

\title{Polarization and charge-separation of moir\'e excitons in van der Waals heterostructures}

\author{Joakim Hagel}
  \affiliation{%
Department of Physics, Chalmers University of Technology, 412 96 Gothenburg, Sweden\\
}%
\author{Samuel Brem}%
\affiliation{%
 Department of Physics, Philipps University of Marburg, 35037 Marburg, Germany\\
}%
  \author{Ermin Malic}%
  \email{ermin.malic@chalmers.se}
  \affiliation{%
 Department of Physics, Philipps University of Marburg, 35037 Marburg, Germany\\
}%
\affiliation{%
Department of Physics, Chalmers University of Technology, 412 96 Gothenburg, Sweden\\
}%

\begin{abstract}
\section*{Abstract}
Twisted transition metal dichalcogenide (TMD) bilayers exhibit periodic moir\'e potentials, which can trap excitons at certain high-symmetry sites. At small twist angles, TMD lattices undergo an atomic reconstruction, altering the moir\'e potential landscape via the formation of large domains, potentially separating the charges in-plane and leading to the formation of intralayer charge-transfer (CT) excitons. Here, we employ a microscopic, material-specific theory to investigate the intralayer charge-separation in atomically reconstructed MoSe$_2$-WSe$_2$ heterostructures. We identify three distinct and twist-angle-dependent exciton regimes including localized Wannier-like excitons, polarized excitons, and intralayer CT excitons. We calculate the moir\'e site hopping for these excitons and predict a fundamentally different twist-angle-dependence compared to regular Wannier excitons - presenting an experimentally accessible key signature for the emergence of intralayer CT excitons. Furthermore, we show that the charge separation and its impact on the hopping can be efficiently tuned via dielectric engineering.
\vfill
{\noindent {\bf Keywords:} van der Waals heterostructures, exciton, moir\'e, charge-transfer}
\end{abstract}
\maketitle

Vertically stacked transition metal dichalcogenides (TMDs) \cite{geim2013van} have revealed intriguing optoelectrical properties \cite{ciarrocchi2022excitonic,mueller2018exciton,splendiani2010emerging,rosati2021dark}, such as strongly bound and long-lived interlayer excitons \cite{rivera2015observation,huang2022spatially,miller2017long,kunstmann2018momentum,deilmann2018interlayer}. Introducing a twist angle between the layers or stacking layers with different lattice constants results in a spatially periodic moir\'e potential, which can trap excitons at certain high-symmetry sites within the moir\'e unit cell \cite{seyler2019signatures,mschmitt2022formation,tran2019evidence,tong2020interferences,yu2017moire,merkl2020twist,brem2020tunable,hagel2022electrical,Forg2021}. Here, exciton transport can be well described in a Bose-Hubbard model via the hopping strength parameter $t$, which describes the tunneling rate of excitons between neighboring moir\'e sites \cite{PhysRevB.105.165419,PhysRevMaterials.6.124002}. The hopping can also be associated with intriguing phenomena, such as superfluidity and Mott insulating phases \cite{PhysRevB.105.165419}, making it an interesting quantity to study for both transport properties and correlated material phases. 

In the regime of small twist angles, the lattice no longer remains rigid, but instead it undergoes a relaxation process known as atomic reconstruction \cite{PhysRevMaterials.8.034001,van2023rotational,weston2020atomic}. In R-type stacked TMDs, this leads to the formation of large triangular domains separated by thin domain walls \cite{weston2020atomic,rosenberger2020twist,van2023rotational,li2021imaging,zhang2017interlayer,Zhao2023,Li2023,sung2020broken,PhysRevMaterials.8.034001,arnold2023relaxation}. The lattice reconstruction considerably changes the moir\'e potential landscape and drastically increases the depth of the moir\'e potential via the induced strain in the material \cite{andersen2021excitons,enaldiev2020stacking,carr2018relaxation,ferreira2021band,PhysRevMaterials.8.034001}. Recent studies have revealed that reconstructed TMD heterostructurescan exhibit varying potential landscapes for electrons and holes, thus separating the particles and consequently forming intralayer charge-transfer (CT) excitons, which can be directly observed via scanning tunneling microscopy \cite{naik2022intralayer,li2024imaging}. However, so far the twist-angle-dependence of the charge separation and its impact on the experimentally accessible hopping has remained in the dark.

\begin{figure}[t!]
\hspace*{-0.5cm}  
\includegraphics[width=1.00\columnwidth]{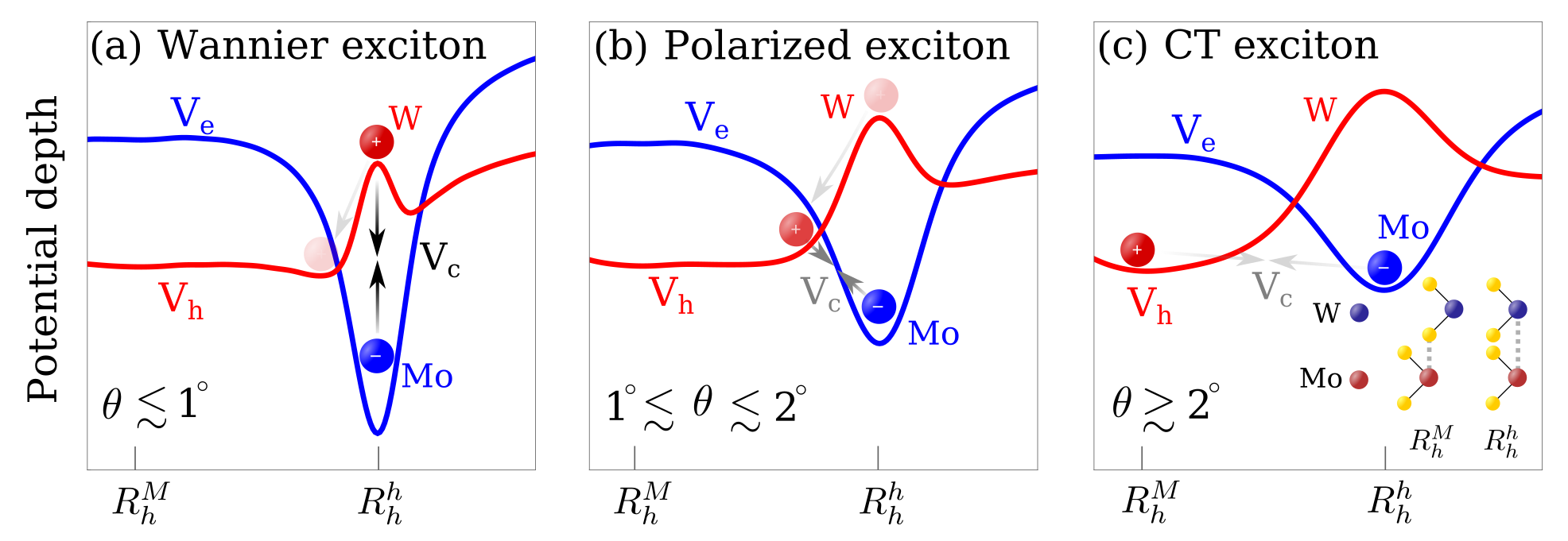}
\caption{\label{fig:1} Schematic showing three different interlayer exciton (electron in Mo and hole in W) regimes in a twisted TMD heterostructure with the respective electron ($V_e$) and hole ($V_h$) potential. \textbf{(a)} In the low twist-angle-regime, electrons and holes are located mostly on top of each other in-plane due to the attractive Coulomb interaction (see arrows for illustration) V$_C$, thus forming localized Wannier-like excitons. \textbf{(b)} In the intermediate twist-angle-regime, there is a slight separation of charges driven by the reconstruction, thus leading to polarized excitons. \textbf{(c)} In the large twist-angle-regime, electrons and holes are centered at different high-symmetry sites in the moir\'e cell leading to the formation of intralayer CT excitons. The inset in (c) shows a schematic for the stacking configurations.}
\end{figure}

In this work, we apply a microscopic and material-specific theory to study the twist-angle-dependence of the in-plane charge separation between electrons and holes in twisted R-type stacked MoSe$_2$-WSe$_2$ heterostructures. We show that by changing the twist angle, the length scale between the electron and hole potential minimums can be efficiently tuned, thus acting as a driving force to separate the charges and counteracting the Coulomb interaction, which attracts electrons and holes to each other. We predict that the competition of the moir\'e potential and the Coulomb interaction leads to three distinct exciton regimes as a function of the changing twist angle: (i) Localized Wannier-like excitons in the small twist-angle range ($\theta<1^\circ$), where electrons and holes are located on top of each other (\autoref{fig:1}.a), (ii) polarized excitons in an intermediate twist-angle range ($1^\circ<\theta<2^\circ$), where holes start becoming separated from electrons within the exciton Bohr radius (\autoref{fig:1}.b), and (iii) intralayer CT excitons in the large twist-angle range ($\theta>2^\circ$), where holes are separated from electrons outside the exciton Bohr radius (\autoref{fig:1}.c). Furthermore, by calculating the inter-site hopping, we show that these three different exciton regimes have qualitatively different twist-angle-dependence compared to regular Wannier excitons. In particular, we predict an unexpected efficient trapping of excitons emerging at larger twist angles. Furthermore, we demonstrate the potential to tune the charge-separation and consequently the hopping rate via dielectric engineering.

\textbf{Theoretical approach:}
In order to model the charge separation between electrons and holes, we develop an approach for the twist-angle-dependent moir\'e potential and the changing length scales between electron and hole minima, which act as the driving force for charge separation. We treat the moir\'e potential $V^{c(v)}(\bm{r})$ in a continuum model as a smoothly varying potential. Taking into account effects of atomic reconstruction, the moir\'e potential is deformed both in depth and geometry \cite{PhysRevMaterials.8.034001}. The atomic displacement giving rise to this deformation is calculated by optimizing the local stacking energy for each twist angle \cite{carr2018relaxation,enaldiev2020stacking}. This allows us to calculate the induced strain in the material and consequently its impact on the electron/hole band structure \cite{PhysRevMaterials.8.034001}. In the R-type stacked MoSe$_2$-WSe$_2$ heterostructure exhibiting a small lattice mismatch, the dominating contribution to the moir\'e potential in the reconstructed regime mainly stems from strain-induced by atomic dilation \cite{van2023rotational}. However, there is also a smaller contribution from the atomic rotation, known as piezo potential \cite{enaldiev2020stacking,Enaldiev_2021}, which is also taken into account in our work. The remaining component is already present in the rigid lattice and is a stacking-dependent polarization-induced shift of the band structure \cite{brem2020tunable}. Here, we focus on KK interlayer excitons as the lowest lying states in the R-type stacked MoSe$_2$-WSe$_2$ heterostructure \cite{PhysRevResearch.3.043217,brem2020tunable}. Since the orbitals around the K-valley mostly consists of d-orbitals around the metal atom, the carrier tunneling is very weak and can thus be neglected \cite{wang2017interlayer,cappelluti2013tight,PhysRevResearch.3.043217}. More details on the modeling of the moir\'e potential and the atomic reconstruction can be found in the supplementary material \cite{supp,khatibi2018impact,rostami2018piezoelectricity}. 

\begin{figure}[t!]
\hspace*{-0.5cm}  
\includegraphics[width=1.00\columnwidth]{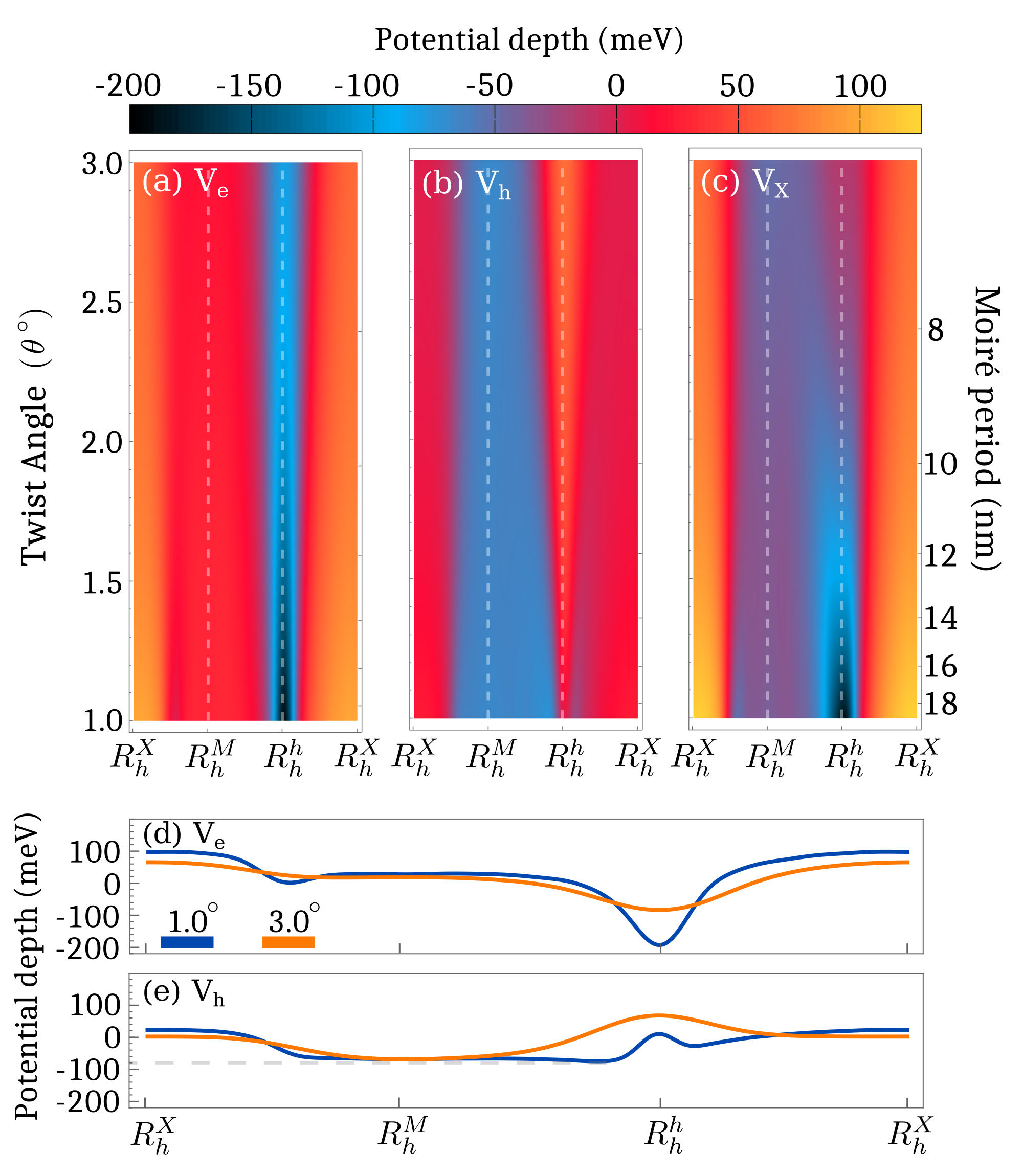}
\caption{\label{fig:2} Moir\'e potential as a function of the twist angle and moir\'e period for the \textbf{(a)} electron, \textbf{(b)} hole, and \textbf{(c)} exciton potential in the R-type stacked MoSe$_2$-WSe$_2$ heterostructure. At small angles, the electron potential is very deep at $R^h_h$ due to the strain induced by atomic reconstruction and becomes more shallow at larger angles. The hole potential is instead positive at $R^h_h$, but has its minimum very close in the $R^M_h$ domain, which is reduced in size with the increasing twist angle. The effective exciton potential $V_X=V_e+V_h$ shows a minimum at $R^h_h$, which changes to $R^M_h$ at larger angles, reflecting the changes in the electron/hole potential. Cut of the \textbf{(d)} electron and \textbf{(e)} hole potential at $\theta=1^{\circ}$, and $\theta=3^{\circ}$. }
\end{figure}

Having calculated the reconstructed moir\'e potential, we can separately plot the contribution of the electron/hole potential as a function of the twist angle and the moir\'e period. In \autoref{fig:2} we show the moir\'e potential as a cut through the supercell for each twist angle, where the coordinates of the high-symmetry stackings ($R^h_h$, $R^M_h$ and $R^X_h$) in the rigid lattice are shown on the x-axis. We find the electron potential to be very deep at $R^h_h$ at small twist angles (up to -200 meV) due to the concentrated strain formed from the atomic reconstruction (\autoref{fig:2}.a) \cite{PhysRevMaterials.8.034001,enaldiev2022self}. The potential then becomes shallower when the lattice starts to revert back to the rigid lattice at larger twist angles. This can also be seen in \autoref{fig:2}.d showing the cut of the electron potential at $\theta=1.0^{\circ}$ and $\theta=3.0^{\circ}$. For the hole potential, the situation is significantly different. Here, the potential has a maximum at $R_h^h$ making it an energetically unfavorable position for a hole, cf. \autoref{fig:2}b. The hole potential instead has its energetic minimum very close to $R^h_h$ due to the increased size of the $R^M_h$ domain and the accumulation of strain around the domain edges (see blue region in \autoref{fig:2}.b). The small distance between the minima (compare minima of blue lines in \autoref{fig:2}.d/e) also translates into the length scale for the separation of charges. At $\theta=1.0^{\circ}$, the distance is about $1$ nm reflecting the narrow width of the $R_h^h$ domains. At larger angles, the moir\'e potential becomes less sharp and the energy minimum is located directly at the $R^M_h$-site instead, cf. \autoref{fig:2}.e, reflecting the diminishing role of the reconstruction. The effective exciton potential $V_X=V_e+V_h$ exhibits a dip at $R^h_h$ (\autoref{fig:2}.c) reflecting the behavior of the electron potential. Increasing the twist angle, it starts to shift towards the hole potential minimum. 

Having access to the twist-angle dependent moir\'e potential in an atomically reconstructed lattice, we can define a Hamilton operator in second quantization, taking into account the attractive Coulomb interaction between electrons and holes. To gain access to exciton wave functions and energies, we derive an eigenvalue equation $H\ket{X}=E\ket{X}$ for the in-plane charge-separated exciton, where $\ket{X}$ is a general exciton state. Introducing the center-of-mass momentum and by employing the zone-folding technique \cite{brem2020tunable,hagel2022electrical,PhysRevMaterials.8.034001}, we can derive the generalized moir\'e exciton eigenvalue equation in a zone-folded basis, which can then be solved by treating it as a sparse matrix to be diagonalized \cite{sanderson2016armadillo,sanderson2019practical}. Further details concerning the generalized moir\'e exciton eigenvalue equation can be found in the the supplemental material \cite{supp}.

\begin{figure}[t!]
\hspace*{-0.5cm}  
\includegraphics[width=1.00\columnwidth]{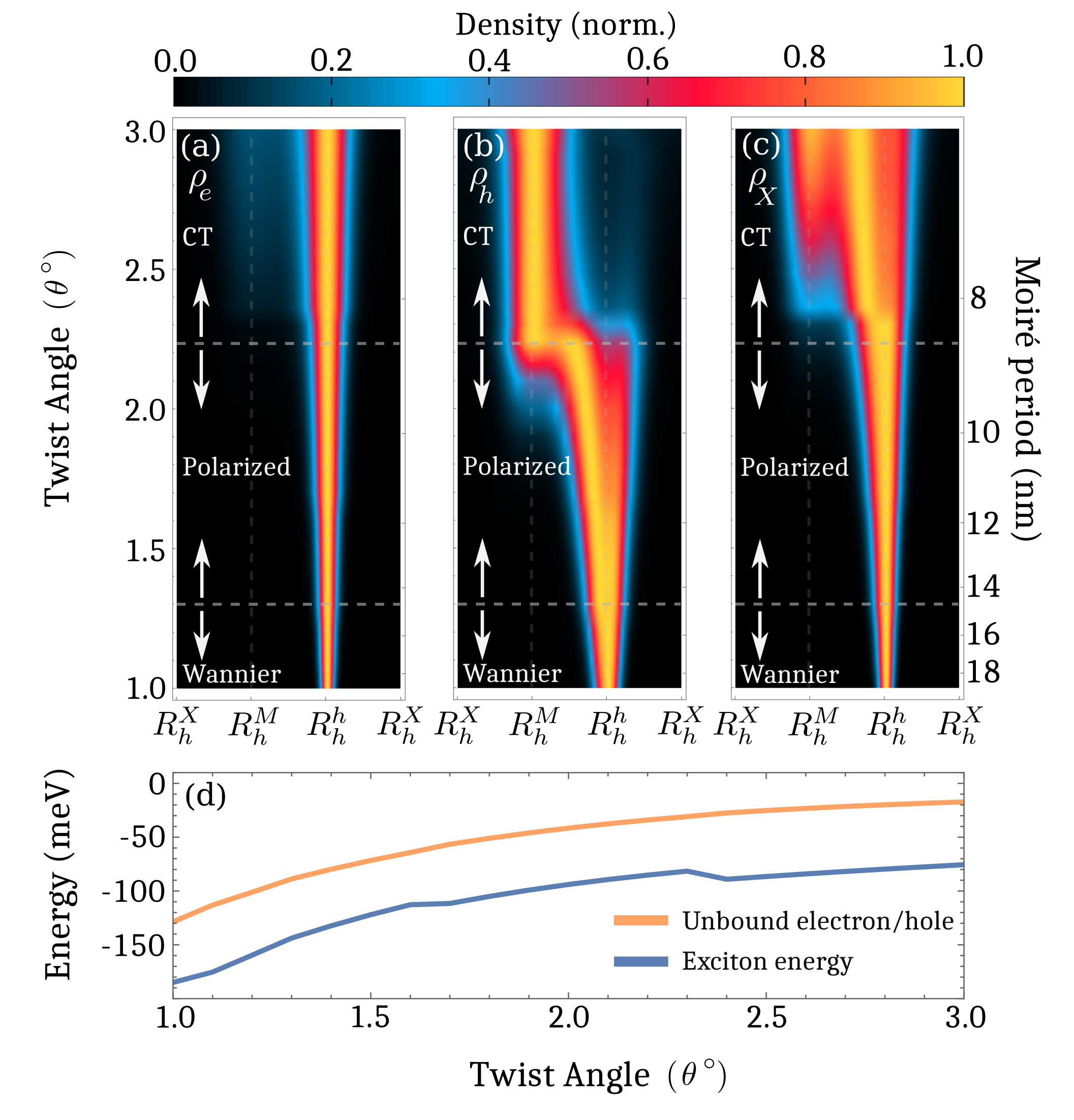}
\caption{\label{fig:3} Electron, hole and exciton densities $\rho_{i}(\bm{r})$ are shown in \textbf{(a-c)}, respectively, as a function of the twist angle and moir\'e period (at $\bm{Q}=\bm{0}$ for excitons) in the R-type stacked MoSe$_2$-WSe$_2$ heterostructure. At small twist angles, the electron density is trapped at $R_h^h$ with holes mostly on top of it, thus constituting a Wannier-like exciton. Increasing the twist angle, pulls the holes away from electrons due to the potential minimum from the reconstructed lattice (cf. also Fig. \ref{fig:1}b), thus forming polarized excitons. At larger twist angles, the impact of the atomic reconstruction becomes negligible and the hole minima now are located directly at $R^M_h$, thus dragging the holes outside of the exciton Bohr radius and forming intralayer CT excitons. \textbf{(d)} Exciton energy (blue) and the unbound electron/hole energy (orange) as a function of twist angle. Here, we can see that the exciton energy is clearly lower and thus constitutes a bound exciton. The energy is given in relation to the free and unbound electron/hole.}
\end{figure}

\textbf{Formation of intralayer CT excitons:}
Diagonalizing the generalized moir\'e exciton eigenvalue problem provides a microscopic access to the two-particle wave function $\Psi_{\bm{k}\bm{k}^{\prime}}$, which can be translated into the electron $\rho_e(\bm{r})$, hole $\rho_h(\bm{r})$ and exciton center-of-mass densities $\rho_X(\bm{r})$, cf. \autoref{fig:3} showing the case of hBN-encapsulated, R-type stacked MoSe$_2$-WSe$_2$ heterostructure. Here, we can see that electrons are very efficiently trapped at low twist angles due to the deep moir\'e potential (\autoref{fig:3}.a). At larger angles, they slowly become more delocalized and start to overlap with the hole trapping location, (cf. the faint blue contribution in the $R^M_h$ region). The hole density overlaps with the electron density at very small twist angles ($\theta\lesssim 1.3^{\circ}$, cf. \autoref{fig:3}.b) thus forming a Wannier-like exciton state, as schematically shown in \autoref{fig:1}.a. In this twist-angle-regime, the length scale between electron and hole minima is such that the strong Coulomb interaction can force holes to be on top of electrons and by just slightly increasing size of the hole density (compare the width of the hole and the electron density in \autoref{fig:3}.a/b) also occupy the hole minimum. This comes as a direct consequence of the proximity of the hole minimum with the narrow electron minimum at $R^h_h$ (compare blue lines in \autoref{fig:2}d/e). Since the Coulomb interaction has pulled holes to be mostly on top of electrons, it can be viewed as a Wannier-like exciton state.

When the twist angle is increased ($1.3^{\circ}\lesssim\theta\lesssim 2.2^{\circ}$), the reconstruction changes and the minimum of the hole potential drags holes away from electrons (cf. the schematic in \autoref{fig:1}.b). Here, the change in the length scale between electron and hole minima gives rise to a second exciton regime, where the Coulomb interaction pulls holes towards electrons that are located in $R^h_h$. In contrast, the moir\'e potential pulls holes further away from $R^h_h$ with increasing twist angle leading to a partial charge separation, cf. \autoref{fig:3}.b in the intermediate twist angle regime. Here, one part of the hole density has been dragged away from the electron density and one part is still overlapping with electrons. However, this charge separation still occurs within the exciton Bohr radius ($\sim 2-3$ nm) and the electron/hole still overlap. Thus, we denote this intermediate twist-angle-regime regime as a polarized exciton state.

In the even larger twist-angle range ($\theta\gtrsim2.2^{\circ}$), the atomic reconstruction has diminished sufficiently and $R^M_h$ domains are nearly reduced to the their size in the rigid lattice moving the hole minimum directly to $R^M_h$ (\autoref{fig:1}.c). Here, holes quickly change the trapping site and are pulled even further away from electrons, now outside of the exciton Bohr radius ($\sim 2-3$ nm). Furthermore, the hole and electron density overlap far less than in the polarized regime, indicating an efficient charge separation. Thus, we denote this state as intralayer CT excitons (cf. the schematic in \autoref{fig:1}.c). Here, the distance between $R^M_h$ and $R_h^h$ is given by $\frac{\sqrt{3}}{3}a_{\text{moir\'e}}$, where $a_{\text{moir\'e}}$ is the moir\'e period (shown on the right y-axis in \autoref{fig:3}). At a supercell size about $8$ nm (corresponding to $\theta\approx 2.4^{\circ}$), we have a charge separation of $\sim 4.6$ nm - in a very good agreement with previous studies on intralayer CT excitons \cite{naik2022intralayer,li2024imaging}. Note that with increasing twist angles, the supercell is decreasing in size and the separation between high-symmetry stackings is itself within the exciton Bohr radius, thus forming polarized excitons again. This is, however, expected to occur above the calculated twist-angle range, where excitons are expected to become entirely delocalized.

The center-of-mass exciton density is shown in \autoref{fig:3}.c. Here, we find the exciton to be very efficiently trapped around $R^h_h$ in the small twist-angle range due to the very deep electron potential (cf. \autoref{fig:2}.a), which is in good agreement with experimental observations \cite{tran2019evidence}. Increasing the twist angle starts to delocalize excitons due to the decreased confinement length and shallower moir\'e potential. When the atomic reconstruction becomes less important and holes becomes trapped at $R^M_h$, the exciton density first becomes more localized, reflecting that its electron/hole constituents are efficiently trapped individually. With larger twist angles, the center-of-mass exciton density again become more delocalized due to the increase in the confinement length. Here, we can see that the exciton density has contributions from both the electron ($R^h_h$) and the hole trapping site ($R^M_h$), but also in the space between them. This reflects that the center-of-mass of the composite electron/hole pair resides in-between the separated charges. Furthermore, \autoref{fig:3}d shows a direct comparison between the exciton energy (blue) and the unbound electron/hole energy (orange) as a function of the twist angle. Here, we observe that both the exciton energy and the unbound electron/hole energy increase with the twist angle, reflecting the change in the width and depth of the moir\'e potential. The exciton energy is clearly lower than the unbound electron/hole energy, indicating that the excitons are still bound. Note that the difference between these energies does not provide the exact binding energy since electrons and holes do not reside in exactly the same location for the two cases. The difference does, however, give a lower bound for the binding energy.

\textbf{Inter-site exciton hopping:} 
Having microscopic access to the twist-angle dependent charge-separated exciton landscape in a reconstructed lattice we can now investigate the moir\'e-site hopping of excitons. By transforming to Wannier basis, the hopping strength $t$ is derived as \cite{PhysRevB.105.165419,PhysRevMaterials.6.124002}
\begin{equation}
\begin{split}
t_{ij}=\frac{1}{N}\sum_{\bm{Q}}e^{i(\bm{R}_i-\bm{R}_j)\cdot\bm{Q}}E_{\bm{Q}},
\end{split}
\end{equation}
where $i(j)$ is the site index, $\bm{R}_i$ is the vector translating between moir\'e sites, and $E_{\bm{Q}}$ is the exciton dispersion as obtained through solving the exciton eigenvalue equation. Considering only the most efficient hopping between the nearest neighboring sites, we calculate the strength of the tunneling between different moir\'e sites as a function of the twist angle.

\begin{figure}[t!]
\hspace*{-0.5cm}  
\includegraphics[width=1.00\columnwidth]{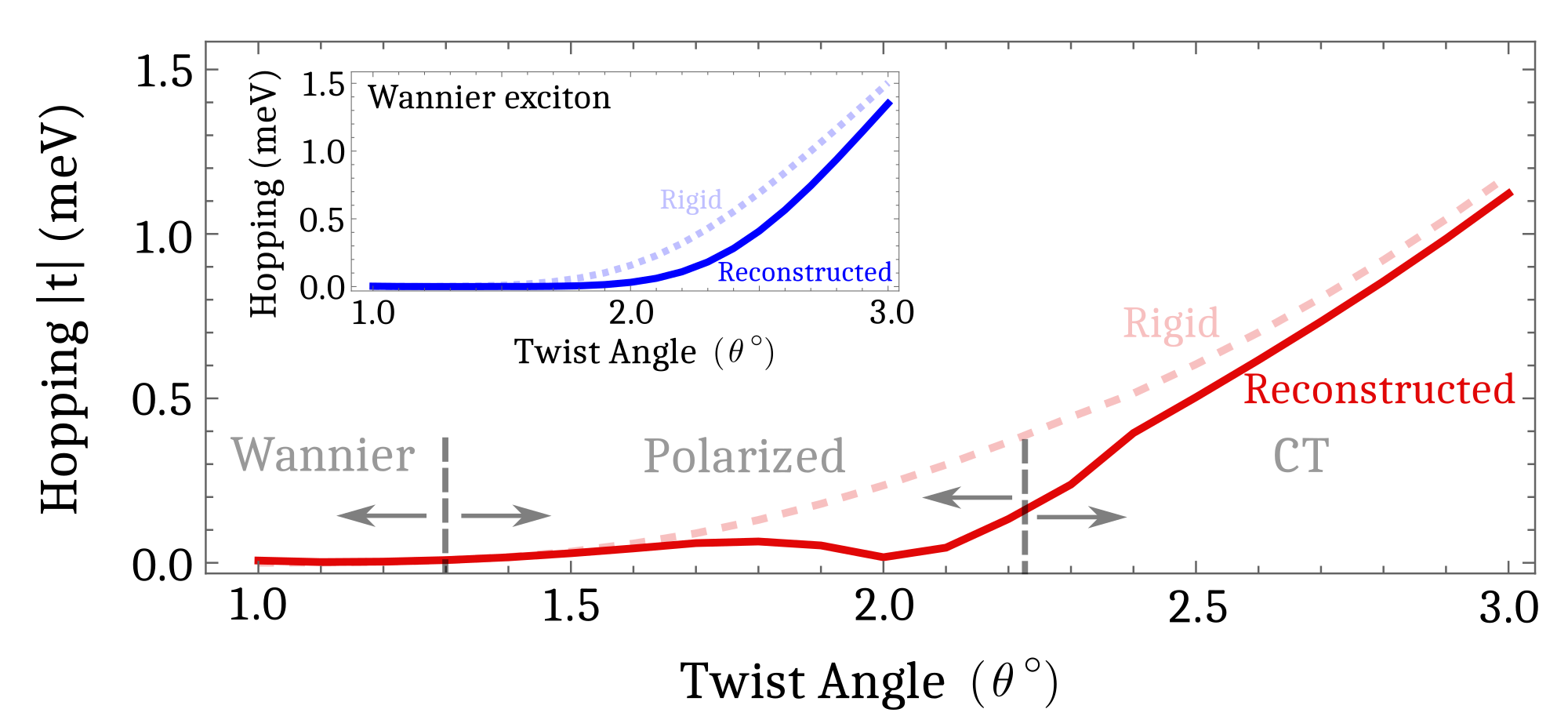}
\caption{\label{fig:4} The hopping strength $|t|$ as a function of the twist angle in the R-type stacked MoSe$_2$-WSe$_2$ heterostructure. In the low-twist-angle range, we find to a large extent non-mobile trapped Wannier-like excitons. Increasing the twist angle, we first observe a noticeable increase in the hopping reflecting the formation of polarized excitons followed by a decrease due to a more efficient separate trapping of electrons and holes. At twist angles larger than 2$^\circ$, the hopping is considerably increased again due to the delocalization of excitons. At $\theta=3.0^{\circ}$ the hopping strength converges to the rigid-lattice case (red dashed line). The inset shows the regular monotonous hopping, when only Wannier excitons are taken into consideration, for both rigid and reconstructed lattice. }
\end{figure}

\autoref{fig:4} shows the hopping strength $|t|$ for hBN-encapsulated MoSe$_2$-WSe$_2$ heterostructure as a function of the twist angle. In the Wannier-like exciton regime ($\theta\lesssim 1.3^{\circ}$) the hopping is negligibly small. When entering the polarized exciton regime ($1.3^{\circ}\lesssim\theta\lesssim 2.2^{\circ}$), the hopping first increases and then goes down again, in stark contrast to the rigid-lattice case (dashed red curve). This unexpected efficient trapping at larger twist angles occurs when electrons and holes starts to become more efficiently trapped individually, i.e the formation of intralayer CT excitons. Increasing the twist angle even further increases the hopping strength and at $\theta=3.0^{\circ}$ it has converged with the rigid-lattice case (dashed red curve), reflecting the diminishing importance of the lattice reconstruction at larger twist angles. Furthermore, the inset shows the hopping when considering only Wannier excitons, i.e no charge-separation is allowed. We find both with and without reconstruction the regular monotonous increase of the hopping strength with the twist angle. This can be traced back to the increased confinement length - in a good agreement with previous studies \cite{PhysRevB.105.165419}. The same behavior is also found for the charge-separated case, but with a rigid lattice. This demonstrates that the emerging unexpected trapping at larger twist angles only occurs in a reconstructed lattice exhibiting a charge separation.

\textbf{Dielectric engineering of exciton hopping:} 
The variation of the twist angle allows us to tune the width of the moir\'e potential and the importance of the atomic reconstruction, and thus the efficiency of the charge separation, as shown in \autoref{fig:3}. Another way to externally influence the charge separation is by tuning the strength of the Coulomb interaction, which can be done by dielectric engineering. So far we have only studied the case of hBN- encapsulated TMD heterostructures. Now we will consider other substrates with lower or higher dielectric screening of the Coulomb potential and investigate the impact on exciton hopping. 

\begin{figure}[t!]
\hspace*{-0.5cm}  
\includegraphics[width=1.00\columnwidth]{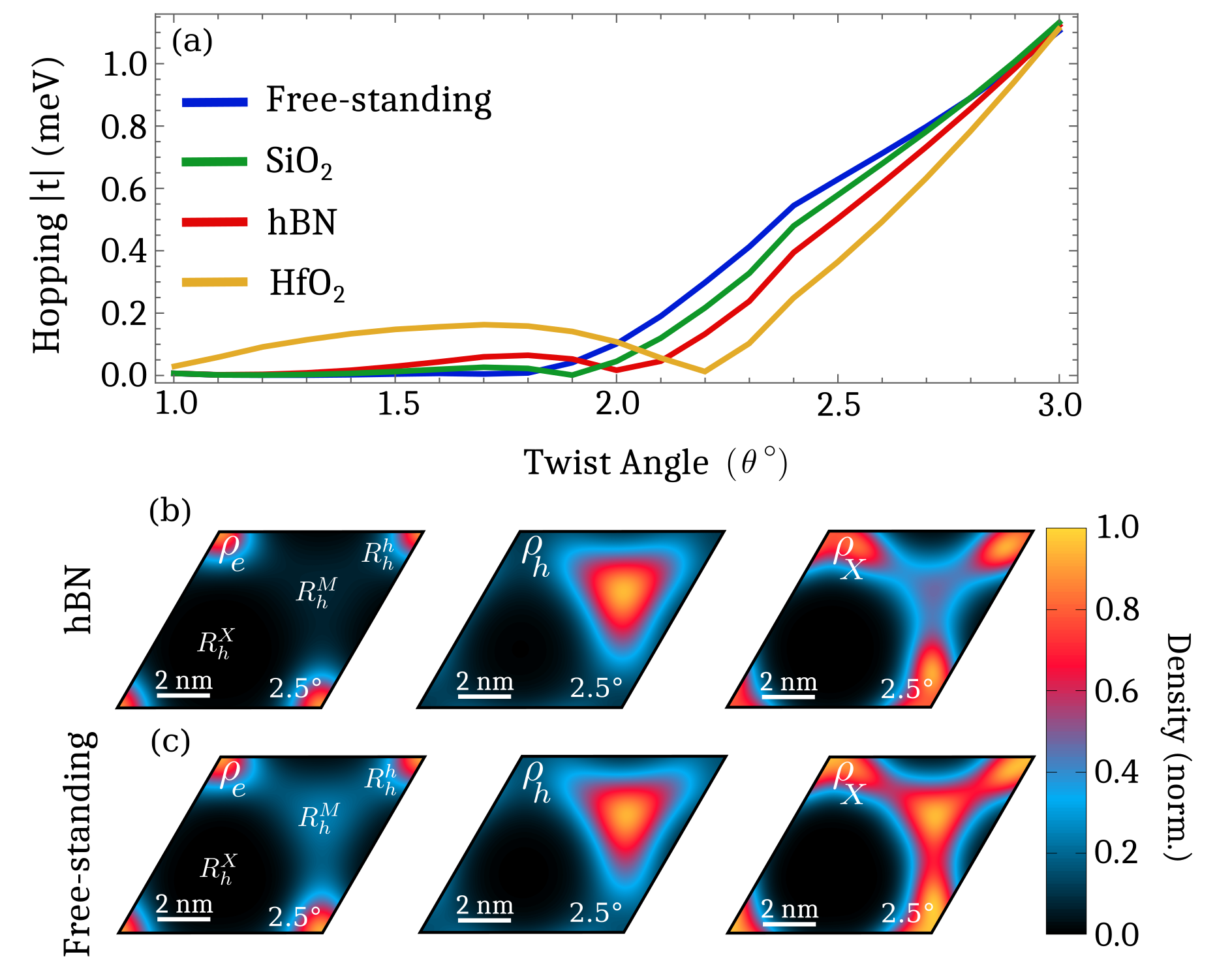}
\caption{\label{fig:5} \textbf{(a)} Exciton hopping strength as a function of the twist angle for different substrates including the free-standing case (blue), SiO$_2$ (green), hBN (red) and HfO$_2$ (yellow), ranging from vanishing screening (with the dielectric constant $\varepsilon=1$) to large screening ($\varepsilon=16.1$) \cite{weng2018honeycomb} in R-type stacked MoSe$_2$-WSe$_2$ heterostructure. The choice of a weaker screening directly impacts the exciton hopping behavior by heavily suppressing the polarized exciton regime due to the stronger Coulomb interaction. The density for electrons, holes and excitons are shown in \textbf{(b)} for the case of hBN-encapsulation and in \textbf{(c)} for the free-standing case at a fixed twist angle of $\theta=2.5^{\circ}$. We observe that electrons have a much larger overlap with holes in the free-standing case, reflecting the stronger Coulomb interaction.
}
\end{figure}

\autoref{fig:5}.a shows the hopping strength $|t|$ for free-standing samples (blue curve) as well as for different substrates including SiO$_2$ (green curve), hBN (red curve) and HfO$_2$ (yellow curve). We find the in the free-standing case, the exciton remains efficiently trapped up to $\theta\approx 1.8^{\circ}$. Here, the increased strength of the Coulomb interaction have heavily suppressed the formation of polarized excitons, in turn forcing holes to be localized on top of electrons instead of being dragged away by the moir\'e potential. In the case of SiO$_2$, a small bump starts to emerge in the intermediate twist-angle range, but the formation of polarized excitons is still very weak. In contrast, considering high-dielectric materials, such as hBN and HfO$_2$ \cite{weng2018honeycomb}, we observe a pronounced increase of the hopping strength in the intermediate twist-angle-regime, which is then reduced when the electron/hole has become efficiently trapped individually. This is emerging due to a significant suppression of the Coulomb interaction resulting in a weaker exciton binding energy and a larger exciton Bohr radius. This allows for holes to become more separated from electrons, while still being inside the Bohr radius, thus enhancing the formation of polarized excitons.

Focusing on a larger twist-angle range, we show that similarly to the hBN case, the hopping for HfO$_2$ is decreased for angles $\theta\gtrsim1.8^{\circ}$, again reflecting the efficient individual trapping of electrons and holes, and the consequential formation of intralayer CT excitons. Interestingly, the high-dielectric substrate reduces the hopping more efficiently in this regime compared to the intermediate twist-angle range, which forms polarized excitons (compare ordering of colors for $\theta\gtrsim2.2^{\circ}$ and $\theta\approx1.8^{\circ}$). This can be understood from the difference in the overlap between electrons and holes. In the regime of intralayer CT excitons, electron and holes are clearly separated outside of the exciton Bohr radius. Decreasing the screening leads to a stronger Coulomb interaction and thus a stronger drive for electrons and holes to be on top of each other, cf. \autoref{fig:5}b/c. This translates into a more delocalized excitons (cf. \autoref{fig:5}.b/c) for a free-standing sample and a larger hopping strength. As a result, we have a qualitatively different behavior in the two different regimes. In the intermediate twist-angle range, excitons are more efficiently trapped by a stronger Coulomb interaction due to the suppression of polarized excitons. In contrast, in the intralayer CT exciton regime, excitons are more efficiently trapped by a weaker Coulomb interaction due to the reduced overlap between electrons and holes. Consequently, we predict that the dielectric engineering of the Coulomb potential can act as an additional external tuning knob to either enhance or suppress exciton trapping.

Furthermore, recent studies have shown that the effective screening can be continuously and dynamically tuned by electrically doping an additional layer that is separated with hBN from the moir\'e structure \cite{tang2022dielectric}. As indicated by our results, this would allow for a continuous control of the charge separation and the Hubbard hopping parameter. Moreover, with the increased separation between electrons and holes, the interaction strength between excitons could also be efficiently tuned \cite{PhysRevB.105.165419}. This is due to the increased exciton dipole length, which would also be of importance when studying high-density moir\'e effects \cite{brem2023bosonic,brem2024optical}.

Our work provides new microscopic insights into the twist-angle-dependent charge separation of moir\'e excitons in an atomically reconstructed lattice. We predict that the interplay between atomic reconstruction and Coulomb interaction leads to three distinct exciton regimes: (i) localized Wannier-like excitons at small twist angles, (ii) polarized excitons in an intermediate twist-angle range, and (iii) charger transfer excitons at larger twist angles. Furthermore, we demonstrate the impact of these three different regimes on moir\'e-site exciton hopping, in particular predicting an unexpected trapping of excitons at larger twist angles. Finally, we predict that dielectric engineering can be used to enhance or suppress the charge separation and its impact on exciton hopping. Overall, our work contributes to a better understanding of moir\'e excitons and the impact of atomic reconstruction on exciton hopping.

\section*{Supporting Information}
The supporting information contains an overview of the theoretical approach used for the atomic reconstruction. Furthermore, it also contains the derivation for the eigenvalue problem and the charge densities.

\section*{Acknowledgments}
This project has received funding from Deutsche Forschungsgemeinschaft via CRC 1083 (project B09) and the regular project 512604469.

\section*{References}


\begin{thebibliography}{51}%
	\makeatletter
	\providecommand \@ifxundefined [1]{%
		\@ifx{#1\undefined}
	}%
	\providecommand \@ifnum [1]{%
		\ifnum #1\expandafter \@firstoftwo
		\else \expandafter \@secondoftwo
		\fi
	}%
	\providecommand \@ifx [1]{%
		\ifx #1\expandafter \@firstoftwo
		\else \expandafter \@secondoftwo
		\fi
	}%
	\providecommand \natexlab [1]{#1}%
	\providecommand \enquote  [1]{``#1''}%
	\providecommand \bibnamefont  [1]{#1}%
	\providecommand \bibfnamefont [1]{#1}%
	\providecommand \citenamefont [1]{#1}%
	\providecommand \href@noop [0]{\@secondoftwo}%
	\providecommand \href [0]{\begingroup \@sanitize@url \@href}%
	\providecommand \@href[1]{\@@startlink{#1}\@@href}%
	\providecommand \@@href[1]{\endgroup#1\@@endlink}%
	\providecommand \@sanitize@url [0]{\catcode `\\12\catcode `\$12\catcode
		`\&12\catcode `\#12\catcode `\^12\catcode `\_12\catcode `\%12\relax}%
	\providecommand \@@startlink[1]{}%
	\providecommand \@@endlink[0]{}%
	\providecommand \url  [0]{\begingroup\@sanitize@url \@url }%
	\providecommand \@url [1]{\endgroup\@href {#1}{\urlprefix }}%
	\providecommand \urlprefix  [0]{URL }%
	\providecommand \Eprint [0]{\href }%
	\providecommand \doibase [0]{https://doi.org/}%
	\providecommand \selectlanguage [0]{\@gobble}%
	\providecommand \bibinfo  [0]{\@secondoftwo}%
	\providecommand \bibfield  [0]{\@secondoftwo}%
	\providecommand \translation [1]{[#1]}%
	\providecommand \BibitemOpen [0]{}%
	\providecommand \bibitemStop [0]{}%
	\providecommand \bibitemNoStop [0]{.\EOS\space}%
	\providecommand \EOS [0]{\spacefactor3000\relax}%
	\providecommand \BibitemShut  [1]{\csname bibitem#1\endcsname}%
	\let\auto@bib@innerbib\@empty
	\bibitem [{\citenamefont {Geim}\ and\ \citenamefont
		{Grigorieva}(2013)}]{geim2013van}%
	\BibitemOpen
	\bibfield  {author} {\bibinfo {author} {\bibfnamefont {A.~K.}\ \bibnamefont
			{Geim}}\ and\ \bibinfo {author} {\bibfnamefont {I.~V.}\ \bibnamefont
			{Grigorieva}},\ }\bibfield  {title} {\bibinfo {title} {Van der waals
			heterostructures},\ }\href@noop {} {\bibfield  {journal} {\bibinfo  {journal}
			{Nature}\ }\textbf {\bibinfo {volume} {499}},\ \bibinfo {pages} {419}
		(\bibinfo {year} {2013})}\BibitemShut {NoStop}%
	\bibitem [{\citenamefont {Ciarrocchi}\ \emph {et~al.}(2022)\citenamefont
		{Ciarrocchi}, \citenamefont {Tagarelli}, \citenamefont {Avsar},\ and\
		\citenamefont {Kis}}]{ciarrocchi2022excitonic}%
	\BibitemOpen
	\bibfield  {author} {\bibinfo {author} {\bibfnamefont {A.}~\bibnamefont
			{Ciarrocchi}}, \bibinfo {author} {\bibfnamefont {F.}~\bibnamefont
			{Tagarelli}}, \bibinfo {author} {\bibfnamefont {A.}~\bibnamefont {Avsar}},\
		and\ \bibinfo {author} {\bibfnamefont {A.}~\bibnamefont {Kis}},\ }\bibfield
	{title} {\bibinfo {title} {Excitonic devices with van der waals
			heterostructures: valleytronics meets twistronics},\ }\href@noop {}
	{\bibfield  {journal} {\bibinfo  {journal} {Nature Reviews Materials}\
		}\textbf {\bibinfo {volume} {7}},\ \bibinfo {pages} {449} (\bibinfo {year}
		{2022})}\BibitemShut {NoStop}%
	\bibitem [{\citenamefont {Mueller}\ and\ \citenamefont
		{Malic}(2018)}]{mueller2018exciton}%
	\BibitemOpen
	\bibfield  {author} {\bibinfo {author} {\bibfnamefont {T.}~\bibnamefont
			{Mueller}}\ and\ \bibinfo {author} {\bibfnamefont {E.}~\bibnamefont
			{Malic}},\ }\bibfield  {title} {\bibinfo {title} {Exciton physics and device
			application of two-dimensional transition metal dichalcogenide
			semiconductors},\ }\href@noop {} {\bibfield  {journal} {\bibinfo  {journal}
			{npj 2D Materials and Applications}\ }\textbf {\bibinfo {volume} {2}},\
		\bibinfo {pages} {1} (\bibinfo {year} {2018})}\BibitemShut {NoStop}%
	\bibitem [{\citenamefont {Splendiani}\ \emph {et~al.}(2010)\citenamefont
		{Splendiani}, \citenamefont {Sun}, \citenamefont {Zhang}, \citenamefont {Li},
		\citenamefont {Kim}, \citenamefont {Chim}, \citenamefont {Galli},\ and\
		\citenamefont {Wang}}]{splendiani2010emerging}%
	\BibitemOpen
	\bibfield  {author} {\bibinfo {author} {\bibfnamefont {A.}~\bibnamefont
			{Splendiani}}, \bibinfo {author} {\bibfnamefont {L.}~\bibnamefont {Sun}},
		\bibinfo {author} {\bibfnamefont {Y.}~\bibnamefont {Zhang}}, \bibinfo
		{author} {\bibfnamefont {T.}~\bibnamefont {Li}}, \bibinfo {author}
		{\bibfnamefont {J.}~\bibnamefont {Kim}}, \bibinfo {author} {\bibfnamefont
			{C.-Y.}\ \bibnamefont {Chim}}, \bibinfo {author} {\bibfnamefont
			{G.}~\bibnamefont {Galli}},\ and\ \bibinfo {author} {\bibfnamefont
			{F.}~\bibnamefont {Wang}},\ }\bibfield  {title} {\bibinfo {title} {Emerging
			photoluminescence in monolayer {MoS$_2$}},\ }\href@noop {} {\bibfield
		{journal} {\bibinfo  {journal} {Nano letters}\ }\textbf {\bibinfo {volume}
			{10}},\ \bibinfo {pages} {1271} (\bibinfo {year} {2010})}\BibitemShut
	{NoStop}%
	\bibitem [{\citenamefont {Rosati}\ \emph {et~al.}(2021)\citenamefont {Rosati},
		\citenamefont {Schmidt}, \citenamefont {Brem}, \citenamefont
		{Perea-Caus{\'\i}n}, \citenamefont {Niehues}, \citenamefont {Kern},
		\citenamefont {Preu{\ss}}, \citenamefont {Schneider}, \citenamefont
		{Michaelis~de Vasconcellos}, \citenamefont {Bratschitsch} \emph
		{et~al.}}]{rosati2021dark}%
	\BibitemOpen
	\bibfield  {author} {\bibinfo {author} {\bibfnamefont {R.}~\bibnamefont
			{Rosati}}, \bibinfo {author} {\bibfnamefont {R.}~\bibnamefont {Schmidt}},
		\bibinfo {author} {\bibfnamefont {S.}~\bibnamefont {Brem}}, \bibinfo {author}
		{\bibfnamefont {R.}~\bibnamefont {Perea-Caus{\'\i}n}}, \bibinfo {author}
		{\bibfnamefont {I.}~\bibnamefont {Niehues}}, \bibinfo {author} {\bibfnamefont
			{J.}~\bibnamefont {Kern}}, \bibinfo {author} {\bibfnamefont {J.~A.}\
			\bibnamefont {Preu{\ss}}}, \bibinfo {author} {\bibfnamefont {R.}~\bibnamefont
			{Schneider}}, \bibinfo {author} {\bibfnamefont {S.}~\bibnamefont
			{Michaelis~de Vasconcellos}}, \bibinfo {author} {\bibfnamefont
			{R.}~\bibnamefont {Bratschitsch}}, \emph {et~al.},\ }\bibfield  {title}
	{\bibinfo {title} {Dark exciton anti-funneling in atomically thin
			semiconductors},\ }\href@noop {} {\bibfield  {journal} {\bibinfo  {journal}
			{Nature Communications}\ }\textbf {\bibinfo {volume} {12}},\ \bibinfo {pages}
		{7221} (\bibinfo {year} {2021})}\BibitemShut {NoStop}%
	\bibitem [{\citenamefont {Rivera}\ \emph {et~al.}(2015)\citenamefont {Rivera},
		\citenamefont {Schaibley}, \citenamefont {Jones}, \citenamefont {Ross},
		\citenamefont {Wu}, \citenamefont {Aivazian}, \citenamefont {Klement},
		\citenamefont {Seyler}, \citenamefont {Clark}, \citenamefont {Ghimire} \emph
		{et~al.}}]{rivera2015observation}%
	\BibitemOpen
	\bibfield  {author} {\bibinfo {author} {\bibfnamefont {P.}~\bibnamefont
			{Rivera}}, \bibinfo {author} {\bibfnamefont {J.~R.}\ \bibnamefont
			{Schaibley}}, \bibinfo {author} {\bibfnamefont {A.~M.}\ \bibnamefont
			{Jones}}, \bibinfo {author} {\bibfnamefont {J.~S.}\ \bibnamefont {Ross}},
		\bibinfo {author} {\bibfnamefont {S.}~\bibnamefont {Wu}}, \bibinfo {author}
		{\bibfnamefont {G.}~\bibnamefont {Aivazian}}, \bibinfo {author}
		{\bibfnamefont {P.}~\bibnamefont {Klement}}, \bibinfo {author} {\bibfnamefont
			{K.}~\bibnamefont {Seyler}}, \bibinfo {author} {\bibfnamefont
			{G.}~\bibnamefont {Clark}}, \bibinfo {author} {\bibfnamefont {N.~J.}\
			\bibnamefont {Ghimire}}, \emph {et~al.},\ }\bibfield  {title} {\bibinfo
		{title} {Observation of long-lived interlayer excitons in monolayer
			{MoSe$_2$}--{WSe$_2$} heterostructures},\ }\href@noop {} {\bibfield
		{journal} {\bibinfo  {journal} {Nature communications}\ }\textbf {\bibinfo
			{volume} {6}},\ \bibinfo {pages} {1} (\bibinfo {year} {2015})}\BibitemShut
	{NoStop}%
	\bibitem [{\citenamefont {Huang}\ \emph {et~al.}(2022)\citenamefont {Huang},
		\citenamefont {Zhao}, \citenamefont {Bo}, \citenamefont {Chu}, \citenamefont
		{Tian}, \citenamefont {Liu}, \citenamefont {Yuan}, \citenamefont {Wu},
		\citenamefont {Zhao}, \citenamefont {Xian} \emph
		{et~al.}}]{huang2022spatially}%
	\BibitemOpen
	\bibfield  {author} {\bibinfo {author} {\bibfnamefont {Z.}~\bibnamefont
			{Huang}}, \bibinfo {author} {\bibfnamefont {Y.}~\bibnamefont {Zhao}},
		\bibinfo {author} {\bibfnamefont {T.}~\bibnamefont {Bo}}, \bibinfo {author}
		{\bibfnamefont {Y.}~\bibnamefont {Chu}}, \bibinfo {author} {\bibfnamefont
			{J.}~\bibnamefont {Tian}}, \bibinfo {author} {\bibfnamefont {L.}~\bibnamefont
			{Liu}}, \bibinfo {author} {\bibfnamefont {Y.}~\bibnamefont {Yuan}}, \bibinfo
		{author} {\bibfnamefont {F.}~\bibnamefont {Wu}}, \bibinfo {author}
		{\bibfnamefont {J.}~\bibnamefont {Zhao}}, \bibinfo {author} {\bibfnamefont
			{L.}~\bibnamefont {Xian}}, \emph {et~al.},\ }\bibfield  {title} {\bibinfo
		{title} {Spatially indirect intervalley excitons in bilayer {WSe$_2$}},\
	}\href@noop {} {\bibfield  {journal} {\bibinfo  {journal} {Physical Review
				B}\ }\textbf {\bibinfo {volume} {105}},\ \bibinfo {pages} {L041409} (\bibinfo
		{year} {2022})}\BibitemShut {NoStop}%
	\bibitem [{\citenamefont {Miller}\ \emph {et~al.}(2017)\citenamefont {Miller},
		\citenamefont {Steinhoff}, \citenamefont {Pano}, \citenamefont {Klein},
		\citenamefont {Jahnke}, \citenamefont {Holleitner},\ and\ \citenamefont
		{Wurstbauer}}]{miller2017long}%
	\BibitemOpen
	\bibfield  {author} {\bibinfo {author} {\bibfnamefont {B.}~\bibnamefont
			{Miller}}, \bibinfo {author} {\bibfnamefont {A.}~\bibnamefont {Steinhoff}},
		\bibinfo {author} {\bibfnamefont {B.}~\bibnamefont {Pano}}, \bibinfo {author}
		{\bibfnamefont {J.}~\bibnamefont {Klein}}, \bibinfo {author} {\bibfnamefont
			{F.}~\bibnamefont {Jahnke}}, \bibinfo {author} {\bibfnamefont
			{A.}~\bibnamefont {Holleitner}},\ and\ \bibinfo {author} {\bibfnamefont
			{U.}~\bibnamefont {Wurstbauer}},\ }\bibfield  {title} {\bibinfo {title}
		{Long-lived direct and indirect interlayer excitons in van der waals
			heterostructures},\ }\href@noop {} {\bibfield  {journal} {\bibinfo  {journal}
			{Nano letters}\ }\textbf {\bibinfo {volume} {17}},\ \bibinfo {pages} {5229}
		(\bibinfo {year} {2017})}\BibitemShut {NoStop}%
	\bibitem [{\citenamefont {Kunstmann}\ \emph {et~al.}(2018)\citenamefont
		{Kunstmann}, \citenamefont {Mooshammer}, \citenamefont {Nagler},
		\citenamefont {Chaves}, \citenamefont {Stein}, \citenamefont {Paradiso},
		\citenamefont {Plechinger}, \citenamefont {Strunk}, \citenamefont
		{Sch{\"u}ller}, \citenamefont {Seifert} \emph
		{et~al.}}]{kunstmann2018momentum}%
	\BibitemOpen
	\bibfield  {author} {\bibinfo {author} {\bibfnamefont {J.}~\bibnamefont
			{Kunstmann}}, \bibinfo {author} {\bibfnamefont {F.}~\bibnamefont
			{Mooshammer}}, \bibinfo {author} {\bibfnamefont {P.}~\bibnamefont {Nagler}},
		\bibinfo {author} {\bibfnamefont {A.}~\bibnamefont {Chaves}}, \bibinfo
		{author} {\bibfnamefont {F.}~\bibnamefont {Stein}}, \bibinfo {author}
		{\bibfnamefont {N.}~\bibnamefont {Paradiso}}, \bibinfo {author}
		{\bibfnamefont {G.}~\bibnamefont {Plechinger}}, \bibinfo {author}
		{\bibfnamefont {C.}~\bibnamefont {Strunk}}, \bibinfo {author} {\bibfnamefont
			{C.}~\bibnamefont {Sch{\"u}ller}}, \bibinfo {author} {\bibfnamefont
			{G.}~\bibnamefont {Seifert}}, \emph {et~al.},\ }\bibfield  {title} {\bibinfo
		{title} {Momentum-space indirect interlayer excitons in transition-metal
			dichalcogenide van der waals heterostructures},\ }\href@noop {} {\bibfield
		{journal} {\bibinfo  {journal} {Nature Physics}\ }\textbf {\bibinfo {volume}
			{14}},\ \bibinfo {pages} {801} (\bibinfo {year} {2018})}\BibitemShut
	{NoStop}%
	\bibitem [{\citenamefont {Deilmann}\ and\ \citenamefont
		{Thygesen}(2018)}]{deilmann2018interlayer}%
	\BibitemOpen
	\bibfield  {author} {\bibinfo {author} {\bibfnamefont {T.}~\bibnamefont
			{Deilmann}}\ and\ \bibinfo {author} {\bibfnamefont {K.~S.}\ \bibnamefont
			{Thygesen}},\ }\bibfield  {title} {\bibinfo {title} {Interlayer excitons with
			large optical amplitudes in layered van der waals materials},\ }\href@noop {}
	{\bibfield  {journal} {\bibinfo  {journal} {Nano letters}\ }\textbf {\bibinfo
			{volume} {18}},\ \bibinfo {pages} {2984} (\bibinfo {year}
		{2018})}\BibitemShut {NoStop}%
	\bibitem [{\citenamefont {Seyler}\ \emph {et~al.}(2019)\citenamefont {Seyler},
		\citenamefont {Rivera}, \citenamefont {Yu}, \citenamefont {Wilson},
		\citenamefont {Ray}, \citenamefont {Mandrus}, \citenamefont {Yan},
		\citenamefont {Yao},\ and\ \citenamefont {Xu}}]{seyler2019signatures}%
	\BibitemOpen
	\bibfield  {author} {\bibinfo {author} {\bibfnamefont {K.~L.}\ \bibnamefont
			{Seyler}}, \bibinfo {author} {\bibfnamefont {P.}~\bibnamefont {Rivera}},
		\bibinfo {author} {\bibfnamefont {H.}~\bibnamefont {Yu}}, \bibinfo {author}
		{\bibfnamefont {N.~P.}\ \bibnamefont {Wilson}}, \bibinfo {author}
		{\bibfnamefont {E.~L.}\ \bibnamefont {Ray}}, \bibinfo {author} {\bibfnamefont
			{D.~G.}\ \bibnamefont {Mandrus}}, \bibinfo {author} {\bibfnamefont
			{J.}~\bibnamefont {Yan}}, \bibinfo {author} {\bibfnamefont {W.}~\bibnamefont
			{Yao}},\ and\ \bibinfo {author} {\bibfnamefont {X.}~\bibnamefont {Xu}},\
	}\bibfield  {title} {\bibinfo {title} {Signatures of moir{\'e}-trapped valley
			excitons in {MoSe2/WSe2} heterobilayers},\ }\href@noop {} {\bibfield
		{journal} {\bibinfo  {journal} {Nature}\ }\textbf {\bibinfo {volume} {567}},\
		\bibinfo {pages} {66} (\bibinfo {year} {2019})}\BibitemShut {NoStop}%
	\bibitem [{\citenamefont {Schmitt}\ \emph {et~al.}(2022)\citenamefont
		{Schmitt}, \citenamefont {Bange}, \citenamefont {Bennecke}, \citenamefont
		{AlMutairi}, \citenamefont {Meneghini}, \citenamefont {Watanabe},
		\citenamefont {Taniguchi}, \citenamefont {Steil}, \citenamefont {Luke},
		\citenamefont {Weitz} \emph {et~al.}}]{mschmitt2022formation}%
	\BibitemOpen
	\bibfield  {author} {\bibinfo {author} {\bibfnamefont {D.}~\bibnamefont
			{Schmitt}}, \bibinfo {author} {\bibfnamefont {J.~P.}\ \bibnamefont {Bange}},
		\bibinfo {author} {\bibfnamefont {W.}~\bibnamefont {Bennecke}}, \bibinfo
		{author} {\bibfnamefont {A.}~\bibnamefont {AlMutairi}}, \bibinfo {author}
		{\bibfnamefont {G.}~\bibnamefont {Meneghini}}, \bibinfo {author}
		{\bibfnamefont {K.}~\bibnamefont {Watanabe}}, \bibinfo {author}
		{\bibfnamefont {T.}~\bibnamefont {Taniguchi}}, \bibinfo {author}
		{\bibfnamefont {D.}~\bibnamefont {Steil}}, \bibinfo {author} {\bibfnamefont
			{D.~R.}\ \bibnamefont {Luke}}, \bibinfo {author} {\bibfnamefont {R.~T.}\
			\bibnamefont {Weitz}}, \emph {et~al.},\ }\bibfield  {title} {\bibinfo {title}
		{Formation of moir{\'e} interlayer excitons in space and time},\ }\href@noop
	{} {\bibfield  {journal} {\bibinfo  {journal} {Nature}\ }\textbf {\bibinfo
			{volume} {608}},\ \bibinfo {pages} {499} (\bibinfo {year}
		{2022})}\BibitemShut {NoStop}%
	\bibitem [{\citenamefont {Tran}\ \emph {et~al.}(2019)\citenamefont {Tran},
		\citenamefont {Moody}, \citenamefont {Wu}, \citenamefont {Lu}, \citenamefont
		{Choi}, \citenamefont {Kim}, \citenamefont {Rai}, \citenamefont {Sanchez},
		\citenamefont {Quan}, \citenamefont {Singh} \emph
		{et~al.}}]{tran2019evidence}%
	\BibitemOpen
	\bibfield  {author} {\bibinfo {author} {\bibfnamefont {K.}~\bibnamefont
			{Tran}}, \bibinfo {author} {\bibfnamefont {G.}~\bibnamefont {Moody}},
		\bibinfo {author} {\bibfnamefont {F.}~\bibnamefont {Wu}}, \bibinfo {author}
		{\bibfnamefont {X.}~\bibnamefont {Lu}}, \bibinfo {author} {\bibfnamefont
			{J.}~\bibnamefont {Choi}}, \bibinfo {author} {\bibfnamefont {K.}~\bibnamefont
			{Kim}}, \bibinfo {author} {\bibfnamefont {A.}~\bibnamefont {Rai}}, \bibinfo
		{author} {\bibfnamefont {D.~A.}\ \bibnamefont {Sanchez}}, \bibinfo {author}
		{\bibfnamefont {J.}~\bibnamefont {Quan}}, \bibinfo {author} {\bibfnamefont
			{A.}~\bibnamefont {Singh}}, \emph {et~al.},\ }\bibfield  {title} {\bibinfo
		{title} {Evidence for moir{\'e} excitons in van der waals heterostructures},\
	}\href@noop {} {\bibfield  {journal} {\bibinfo  {journal} {Nature}\ }\textbf
		{\bibinfo {volume} {567}},\ \bibinfo {pages} {71} (\bibinfo {year}
		{2019})}\BibitemShut {NoStop}%
	\bibitem [{\citenamefont {Tong}\ \emph {et~al.}(2020)\citenamefont {Tong},
		\citenamefont {Chen}, \citenamefont {Xiao}, \citenamefont {Yu},\ and\
		\citenamefont {Yao}}]{tong2020interferences}%
	\BibitemOpen
	\bibfield  {author} {\bibinfo {author} {\bibfnamefont {Q.}~\bibnamefont
			{Tong}}, \bibinfo {author} {\bibfnamefont {M.}~\bibnamefont {Chen}}, \bibinfo
		{author} {\bibfnamefont {F.}~\bibnamefont {Xiao}}, \bibinfo {author}
		{\bibfnamefont {H.}~\bibnamefont {Yu}},\ and\ \bibinfo {author}
		{\bibfnamefont {W.}~\bibnamefont {Yao}},\ }\bibfield  {title} {\bibinfo
		{title} {Interferences of electrostatic moir{\'e} potentials and bichromatic
			superlattices of electrons and excitons in transition metal
			dichalcogenides},\ }\href@noop {} {\bibfield  {journal} {\bibinfo  {journal}
			{2D Materials}\ }\textbf {\bibinfo {volume} {8}},\ \bibinfo {pages} {025007}
		(\bibinfo {year} {2020})}\BibitemShut {NoStop}%
	\bibitem [{\citenamefont {Yu}\ \emph {et~al.}(2017)\citenamefont {Yu},
		\citenamefont {Liu}, \citenamefont {Tang}, \citenamefont {Xu},\ and\
		\citenamefont {Yao}}]{yu2017moire}%
	\BibitemOpen
	\bibfield  {author} {\bibinfo {author} {\bibfnamefont {H.}~\bibnamefont
			{Yu}}, \bibinfo {author} {\bibfnamefont {G.-B.}\ \bibnamefont {Liu}},
		\bibinfo {author} {\bibfnamefont {J.}~\bibnamefont {Tang}}, \bibinfo {author}
		{\bibfnamefont {X.}~\bibnamefont {Xu}},\ and\ \bibinfo {author}
		{\bibfnamefont {W.}~\bibnamefont {Yao}},\ }\bibfield  {title} {\bibinfo
		{title} {Moir{\'e} excitons: From programmable quantum emitter arrays to
			spin-orbit--coupled artificial lattices},\ }\href@noop {} {\bibfield
		{journal} {\bibinfo  {journal} {Science advances}\ }\textbf {\bibinfo
			{volume} {3}},\ \bibinfo {pages} {e1701696} (\bibinfo {year}
		{2017})}\BibitemShut {NoStop}%
	\bibitem [{\citenamefont {Merkl}\ \emph {et~al.}(2020)\citenamefont {Merkl},
		\citenamefont {Mooshammer}, \citenamefont {Brem}, \citenamefont {Girnghuber},
		\citenamefont {Lin}, \citenamefont {Weigl}, \citenamefont {Liebich},
		\citenamefont {Yong}, \citenamefont {Gillen}, \citenamefont {Maultzsch} \emph
		{et~al.}}]{merkl2020twist}%
	\BibitemOpen
	\bibfield  {author} {\bibinfo {author} {\bibfnamefont {P.}~\bibnamefont
			{Merkl}}, \bibinfo {author} {\bibfnamefont {F.}~\bibnamefont {Mooshammer}},
		\bibinfo {author} {\bibfnamefont {S.}~\bibnamefont {Brem}}, \bibinfo {author}
		{\bibfnamefont {A.}~\bibnamefont {Girnghuber}}, \bibinfo {author}
		{\bibfnamefont {K.-Q.}\ \bibnamefont {Lin}}, \bibinfo {author} {\bibfnamefont
			{L.}~\bibnamefont {Weigl}}, \bibinfo {author} {\bibfnamefont
			{M.}~\bibnamefont {Liebich}}, \bibinfo {author} {\bibfnamefont {C.-K.}\
			\bibnamefont {Yong}}, \bibinfo {author} {\bibfnamefont {R.}~\bibnamefont
			{Gillen}}, \bibinfo {author} {\bibfnamefont {J.}~\bibnamefont {Maultzsch}},
		\emph {et~al.},\ }\bibfield  {title} {\bibinfo {title} {Twist-tailoring
			coulomb correlations in van der waals homobilayers},\ }\href@noop {}
	{\bibfield  {journal} {\bibinfo  {journal} {Nature communications}\ }\textbf
		{\bibinfo {volume} {11}},\ \bibinfo {pages} {1} (\bibinfo {year}
		{2020})}\BibitemShut {NoStop}%
	\bibitem [{\citenamefont {Brem}\ \emph {et~al.}(2020)\citenamefont {Brem},
		\citenamefont {Linder\"alv}, \citenamefont {Erhart},\ and\ \citenamefont
		{Malic}}]{brem2020tunable}%
	\BibitemOpen
	\bibfield  {author} {\bibinfo {author} {\bibfnamefont {S.}~\bibnamefont
			{Brem}}, \bibinfo {author} {\bibfnamefont {C.}~\bibnamefont {Linder\"alv}},
		\bibinfo {author} {\bibfnamefont {P.}~\bibnamefont {Erhart}},\ and\ \bibinfo
		{author} {\bibfnamefont {E.}~\bibnamefont {Malic}},\ }\bibfield  {title}
	{\bibinfo {title} {Tunable phases of moiré excitons in van der waals
			heterostructures},\ }\href {https://doi.org/10.1021/acs.nanolett.0c03019}
	{\bibfield  {journal} {\bibinfo  {journal} {Nano Letters}\ }\textbf {\bibinfo
			{volume} {20}},\ \bibinfo {pages} {8534} (\bibinfo {year} {2020})},\ \bibinfo
	{note} {pMID: 32970445},\ \Eprint
	{https://arxiv.org/abs/https://doi.org/10.1021/acs.nanolett.0c03019}
	{https://doi.org/10.1021/acs.nanolett.0c03019} \BibitemShut {NoStop}%
	\bibitem [{\citenamefont {Hagel}\ \emph {et~al.}(2023)\citenamefont {Hagel},
		\citenamefont {Brem},\ and\ \citenamefont {Malic}}]{hagel2022electrical}%
	\BibitemOpen
	\bibfield  {author} {\bibinfo {author} {\bibfnamefont {J.}~\bibnamefont
			{Hagel}}, \bibinfo {author} {\bibfnamefont {S.}~\bibnamefont {Brem}},\ and\
		\bibinfo {author} {\bibfnamefont {E.}~\bibnamefont {Malic}},\ }\bibfield
	{title} {\bibinfo {title} {Electrical tuning of moir{\'e} excitons in
			{MoSe$_2$} bilayers},\ }\href@noop {} {\bibfield  {journal} {\bibinfo
			{journal} {2D Materials}\ }\textbf {\bibinfo {volume} {10}},\ \bibinfo
		{pages} {014013} (\bibinfo {year} {2023})}\BibitemShut {NoStop}%
	\bibitem [{\citenamefont {F{\"o}rg}\ \emph {et~al.}(2021)\citenamefont
		{F{\"o}rg}, \citenamefont {Baimuratov}, \citenamefont {Kruchinin},
		\citenamefont {Vovk}, \citenamefont {Scherzer}, \citenamefont {F{\"o}rste},
		\citenamefont {Funk}, \citenamefont {Watanabe}, \citenamefont {Taniguchi},\
		and\ \citenamefont {H{\"o}gele}}]{Forg2021}%
	\BibitemOpen
	\bibfield  {author} {\bibinfo {author} {\bibfnamefont {M.}~\bibnamefont
			{F{\"o}rg}}, \bibinfo {author} {\bibfnamefont {A.~S.}\ \bibnamefont
			{Baimuratov}}, \bibinfo {author} {\bibfnamefont {S.~Y.}\ \bibnamefont
			{Kruchinin}}, \bibinfo {author} {\bibfnamefont {I.~A.}\ \bibnamefont {Vovk}},
		\bibinfo {author} {\bibfnamefont {J.}~\bibnamefont {Scherzer}}, \bibinfo
		{author} {\bibfnamefont {J.}~\bibnamefont {F{\"o}rste}}, \bibinfo {author}
		{\bibfnamefont {V.}~\bibnamefont {Funk}}, \bibinfo {author} {\bibfnamefont
			{K.}~\bibnamefont {Watanabe}}, \bibinfo {author} {\bibfnamefont
			{T.}~\bibnamefont {Taniguchi}},\ and\ \bibinfo {author} {\bibfnamefont
			{A.}~\bibnamefont {H{\"o}gele}},\ }\bibfield  {title} {\bibinfo {title}
		{Moir{\'e} excitons in {MoSe2-WSe2} heterobilayers and heterotrilayers},\
	}\href {https://doi.org/10.1038/s41467-021-21822-z} {\bibfield  {journal}
		{\bibinfo  {journal} {Nature Communications}\ }\textbf {\bibinfo {volume}
			{12}},\ \bibinfo {pages} {1656} (\bibinfo {year} {2021})}\BibitemShut
	{NoStop}%
	\bibitem [{\citenamefont {G\"otting}\ \emph {et~al.}(2022)\citenamefont
		{G\"otting}, \citenamefont {Lohof},\ and\ \citenamefont
		{Gies}}]{PhysRevB.105.165419}%
	\BibitemOpen
	\bibfield  {author} {\bibinfo {author} {\bibfnamefont {N.}~\bibnamefont
			{G\"otting}}, \bibinfo {author} {\bibfnamefont {F.}~\bibnamefont {Lohof}},\
		and\ \bibinfo {author} {\bibfnamefont {C.}~\bibnamefont {Gies}},\ }\bibfield
	{title} {\bibinfo {title} {Moir\'e-bose-hubbard model for interlayer excitons
			in twisted transition metal dichalcogenide heterostructures},\ }\href
	{https://doi.org/10.1103/PhysRevB.105.165419} {\bibfield  {journal} {\bibinfo
			{journal} {Phys. Rev. B}\ }\textbf {\bibinfo {volume} {105}},\ \bibinfo
		{pages} {165419} (\bibinfo {year} {2022})}\BibitemShut {NoStop}%
	\bibitem [{\citenamefont {Knorr}\ \emph {et~al.}(2022)\citenamefont {Knorr},
		\citenamefont {Brem}, \citenamefont {Meneghini},\ and\ \citenamefont
		{Malic}}]{PhysRevMaterials.6.124002}%
	\BibitemOpen
	\bibfield  {author} {\bibinfo {author} {\bibfnamefont {W.}~\bibnamefont
			{Knorr}}, \bibinfo {author} {\bibfnamefont {S.}~\bibnamefont {Brem}},
		\bibinfo {author} {\bibfnamefont {G.}~\bibnamefont {Meneghini}},\ and\
		\bibinfo {author} {\bibfnamefont {E.}~\bibnamefont {Malic}},\ }\bibfield
	{title} {\bibinfo {title} {Exciton transport in a moir\'e potential: From
			hopping to dispersive regime},\ }\href
	{https://doi.org/10.1103/PhysRevMaterials.6.124002} {\bibfield  {journal}
		{\bibinfo  {journal} {Phys. Rev. Mater.}\ }\textbf {\bibinfo {volume} {6}},\
		\bibinfo {pages} {124002} (\bibinfo {year} {2022})}\BibitemShut {NoStop}%
	\bibitem [{\citenamefont {Hagel}\ \emph {et~al.}(2024)\citenamefont {Hagel},
		\citenamefont {Brem}, \citenamefont {Pineiro},\ and\ \citenamefont
		{Malic}}]{PhysRevMaterials.8.034001}%
	\BibitemOpen
	\bibfield  {author} {\bibinfo {author} {\bibfnamefont {J.}~\bibnamefont
			{Hagel}}, \bibinfo {author} {\bibfnamefont {S.}~\bibnamefont {Brem}},
		\bibinfo {author} {\bibfnamefont {J.~A.}\ \bibnamefont {Pineiro}},\ and\
		\bibinfo {author} {\bibfnamefont {E.}~\bibnamefont {Malic}},\ }\bibfield
	{title} {\bibinfo {title} {Impact of atomic reconstruction on optical spectra
			of twisted tmd homobilayers},\ }\href
	{https://doi.org/10.1103/PhysRevMaterials.8.034001} {\bibfield  {journal}
		{\bibinfo  {journal} {Phys. Rev. Mater.}\ }\textbf {\bibinfo {volume} {8}},\
		\bibinfo {pages} {034001} (\bibinfo {year} {2024})}\BibitemShut {NoStop}%
	\bibitem [{\citenamefont {Van~Winkle}\ \emph {et~al.}(2023)\citenamefont
		{Van~Winkle}, \citenamefont {Craig}, \citenamefont {Carr}, \citenamefont
		{Dandu}, \citenamefont {Bustillo}, \citenamefont {Ciston}, \citenamefont
		{Ophus}, \citenamefont {Taniguchi}, \citenamefont {Watanabe}, \citenamefont
		{Raja} \emph {et~al.}}]{van2023rotational}%
	\BibitemOpen
	\bibfield  {author} {\bibinfo {author} {\bibfnamefont {M.}~\bibnamefont
			{Van~Winkle}}, \bibinfo {author} {\bibfnamefont {I.~M.}\ \bibnamefont
			{Craig}}, \bibinfo {author} {\bibfnamefont {S.}~\bibnamefont {Carr}},
		\bibinfo {author} {\bibfnamefont {M.}~\bibnamefont {Dandu}}, \bibinfo
		{author} {\bibfnamefont {K.~C.}\ \bibnamefont {Bustillo}}, \bibinfo {author}
		{\bibfnamefont {J.}~\bibnamefont {Ciston}}, \bibinfo {author} {\bibfnamefont
			{C.}~\bibnamefont {Ophus}}, \bibinfo {author} {\bibfnamefont
			{T.}~\bibnamefont {Taniguchi}}, \bibinfo {author} {\bibfnamefont
			{K.}~\bibnamefont {Watanabe}}, \bibinfo {author} {\bibfnamefont
			{A.}~\bibnamefont {Raja}}, \emph {et~al.},\ }\bibfield  {title} {\bibinfo
		{title} {Rotational and dilational reconstruction in transition metal
			dichalcogenide moir{\'e} bilayers},\ }\href@noop {} {\bibfield  {journal}
		{\bibinfo  {journal} {Nature Communications}\ }\textbf {\bibinfo {volume}
			{14}},\ \bibinfo {pages} {2989} (\bibinfo {year} {2023})}\BibitemShut
	{NoStop}%
	\bibitem [{\citenamefont {Weston}\ \emph {et~al.}(2020)\citenamefont {Weston},
		\citenamefont {Zou}, \citenamefont {Enaldiev}, \citenamefont {Summerfield},
		\citenamefont {Clark}, \citenamefont {Z{\'o}lyomi}, \citenamefont {Graham},
		\citenamefont {Yelgel}, \citenamefont {Magorrian}, \citenamefont {Zhou} \emph
		{et~al.}}]{weston2020atomic}%
	\BibitemOpen
	\bibfield  {author} {\bibinfo {author} {\bibfnamefont {A.}~\bibnamefont
			{Weston}}, \bibinfo {author} {\bibfnamefont {Y.}~\bibnamefont {Zou}},
		\bibinfo {author} {\bibfnamefont {V.}~\bibnamefont {Enaldiev}}, \bibinfo
		{author} {\bibfnamefont {A.}~\bibnamefont {Summerfield}}, \bibinfo {author}
		{\bibfnamefont {N.}~\bibnamefont {Clark}}, \bibinfo {author} {\bibfnamefont
			{V.}~\bibnamefont {Z{\'o}lyomi}}, \bibinfo {author} {\bibfnamefont
			{A.}~\bibnamefont {Graham}}, \bibinfo {author} {\bibfnamefont
			{C.}~\bibnamefont {Yelgel}}, \bibinfo {author} {\bibfnamefont
			{S.}~\bibnamefont {Magorrian}}, \bibinfo {author} {\bibfnamefont
			{M.}~\bibnamefont {Zhou}}, \emph {et~al.},\ }\bibfield  {title} {\bibinfo
		{title} {Atomic reconstruction in twisted bilayers of transition metal
			dichalcogenides},\ }\href@noop {} {\bibfield  {journal} {\bibinfo  {journal}
			{Nature Nanotechnology}\ }\textbf {\bibinfo {volume} {15}},\ \bibinfo {pages}
		{592} (\bibinfo {year} {2020})}\BibitemShut {NoStop}%
	\bibitem [{\citenamefont {Rosenberger}\ \emph {et~al.}(2020)\citenamefont
		{Rosenberger}, \citenamefont {Chuang}, \citenamefont {Phillips},
		\citenamefont {Oleshko}, \citenamefont {McCreary}, \citenamefont {Sivaram},
		\citenamefont {Hellberg},\ and\ \citenamefont
		{Jonker}}]{rosenberger2020twist}%
	\BibitemOpen
	\bibfield  {author} {\bibinfo {author} {\bibfnamefont {M.~R.}\ \bibnamefont
			{Rosenberger}}, \bibinfo {author} {\bibfnamefont {H.-J.}\ \bibnamefont
			{Chuang}}, \bibinfo {author} {\bibfnamefont {M.}~\bibnamefont {Phillips}},
		\bibinfo {author} {\bibfnamefont {V.~P.}\ \bibnamefont {Oleshko}}, \bibinfo
		{author} {\bibfnamefont {K.~M.}\ \bibnamefont {McCreary}}, \bibinfo {author}
		{\bibfnamefont {S.~V.}\ \bibnamefont {Sivaram}}, \bibinfo {author}
		{\bibfnamefont {C.~S.}\ \bibnamefont {Hellberg}},\ and\ \bibinfo {author}
		{\bibfnamefont {B.~T.}\ \bibnamefont {Jonker}},\ }\bibfield  {title}
	{\bibinfo {title} {Twist angle-dependent atomic reconstruction and moir{\'e}
			patterns in transition metal dichalcogenide heterostructures},\ }\href@noop
	{} {\bibfield  {journal} {\bibinfo  {journal} {ACS nano}\ }\textbf {\bibinfo
			{volume} {14}},\ \bibinfo {pages} {4550} (\bibinfo {year}
		{2020})}\BibitemShut {NoStop}%
	\bibitem [{\citenamefont {Li}\ \emph {et~al.}(2021)\citenamefont {Li},
		\citenamefont {Li}, \citenamefont {Naik}, \citenamefont {Xie}, \citenamefont
		{Li}, \citenamefont {Wang}, \citenamefont {Regan}, \citenamefont {Wang},
		\citenamefont {Zhao}, \citenamefont {Zhao} \emph {et~al.}}]{li2021imaging}%
	\BibitemOpen
	\bibfield  {author} {\bibinfo {author} {\bibfnamefont {H.}~\bibnamefont
			{Li}}, \bibinfo {author} {\bibfnamefont {S.}~\bibnamefont {Li}}, \bibinfo
		{author} {\bibfnamefont {M.~H.}\ \bibnamefont {Naik}}, \bibinfo {author}
		{\bibfnamefont {J.}~\bibnamefont {Xie}}, \bibinfo {author} {\bibfnamefont
			{X.}~\bibnamefont {Li}}, \bibinfo {author} {\bibfnamefont {J.}~\bibnamefont
			{Wang}}, \bibinfo {author} {\bibfnamefont {E.}~\bibnamefont {Regan}},
		\bibinfo {author} {\bibfnamefont {D.}~\bibnamefont {Wang}}, \bibinfo {author}
		{\bibfnamefont {W.}~\bibnamefont {Zhao}}, \bibinfo {author} {\bibfnamefont
			{S.}~\bibnamefont {Zhao}}, \emph {et~al.},\ }\bibfield  {title} {\bibinfo
		{title} {Imaging moir{\'e} flat bands in three-dimensional reconstructed
			{WSe2/WS2} superlattices},\ }\href@noop {} {\bibfield  {journal} {\bibinfo
			{journal} {Nature materials}\ }\textbf {\bibinfo {volume} {20}},\ \bibinfo
		{pages} {945} (\bibinfo {year} {2021})}\BibitemShut {NoStop}%
	\bibitem [{\citenamefont {Zhang}\ \emph {et~al.}(2017)\citenamefont {Zhang},
		\citenamefont {Chuu}, \citenamefont {Ren}, \citenamefont {Li}, \citenamefont
		{Li}, \citenamefont {Jin}, \citenamefont {Chou},\ and\ \citenamefont
		{Shih}}]{zhang2017interlayer}%
	\BibitemOpen
	\bibfield  {author} {\bibinfo {author} {\bibfnamefont {C.}~\bibnamefont
			{Zhang}}, \bibinfo {author} {\bibfnamefont {C.-P.}\ \bibnamefont {Chuu}},
		\bibinfo {author} {\bibfnamefont {X.}~\bibnamefont {Ren}}, \bibinfo {author}
		{\bibfnamefont {M.-Y.}\ \bibnamefont {Li}}, \bibinfo {author} {\bibfnamefont
			{L.-J.}\ \bibnamefont {Li}}, \bibinfo {author} {\bibfnamefont
			{C.}~\bibnamefont {Jin}}, \bibinfo {author} {\bibfnamefont {M.-Y.}\
			\bibnamefont {Chou}},\ and\ \bibinfo {author} {\bibfnamefont {C.-K.}\
			\bibnamefont {Shih}},\ }\bibfield  {title} {\bibinfo {title} {Interlayer
			couplings, moir{\'e} patterns, and 2d electronic superlattices in {MoS2/WSe2}
			hetero-bilayers},\ }\href@noop {} {\bibfield  {journal} {\bibinfo  {journal}
			{Science advances}\ }\textbf {\bibinfo {volume} {3}},\ \bibinfo {pages}
		{e1601459} (\bibinfo {year} {2017})}\BibitemShut {NoStop}%
	\bibitem [{\citenamefont {Zhao}\ \emph {et~al.}(2023)\citenamefont {Zhao},
		\citenamefont {Li}, \citenamefont {Huang}, \citenamefont {Rupp},
		\citenamefont {G{\"o}ser}, \citenamefont {Vovk}, \citenamefont {Kruchinin},
		\citenamefont {Watanabe}, \citenamefont {Taniguchi}, \citenamefont {Bilgin},
		\citenamefont {Baimuratov},\ and\ \citenamefont {H{\"o}gele}}]{Zhao2023}%
	\BibitemOpen
	\bibfield  {author} {\bibinfo {author} {\bibfnamefont {S.}~\bibnamefont
			{Zhao}}, \bibinfo {author} {\bibfnamefont {Z.}~\bibnamefont {Li}}, \bibinfo
		{author} {\bibfnamefont {X.}~\bibnamefont {Huang}}, \bibinfo {author}
		{\bibfnamefont {A.}~\bibnamefont {Rupp}}, \bibinfo {author} {\bibfnamefont
			{J.}~\bibnamefont {G{\"o}ser}}, \bibinfo {author} {\bibfnamefont {I.~A.}\
			\bibnamefont {Vovk}}, \bibinfo {author} {\bibfnamefont {S.~Y.}\ \bibnamefont
			{Kruchinin}}, \bibinfo {author} {\bibfnamefont {K.}~\bibnamefont {Watanabe}},
		\bibinfo {author} {\bibfnamefont {T.}~\bibnamefont {Taniguchi}}, \bibinfo
		{author} {\bibfnamefont {I.}~\bibnamefont {Bilgin}}, \bibinfo {author}
		{\bibfnamefont {A.~S.}\ \bibnamefont {Baimuratov}},\ and\ \bibinfo {author}
		{\bibfnamefont {A.}~\bibnamefont {H{\"o}gele}},\ }\bibfield  {title}
	{\bibinfo {title} {Excitons in mesoscopically reconstructed moir{\'e}
			heterostructures},\ }\href {https://doi.org/10.1038/s41565-023-01356-9}
	{\bibfield  {journal} {\bibinfo  {journal} {Nature Nanotechnology}\ }\textbf
		{\bibinfo {volume} {18}},\ \bibinfo {pages} {572} (\bibinfo {year}
		{2023})}\BibitemShut {NoStop}%
	\bibitem [{\citenamefont {Li}\ \emph {et~al.}(2023)\citenamefont {Li},
		\citenamefont {Tabataba-Vakili}, \citenamefont {Zhao}, \citenamefont {Rupp},
		\citenamefont {Bilgin}, \citenamefont {Herdegen}, \citenamefont {M{\"a}rz},
		\citenamefont {Watanabe}, \citenamefont {Taniguchi}, \citenamefont
		{Schleder}, \citenamefont {Baimuratov}, \citenamefont {Kaxiras},
		\citenamefont {M{\"u}ller-Caspary},\ and\ \citenamefont
		{H{\"o}gele}}]{Li2023}%
	\BibitemOpen
	\bibfield  {author} {\bibinfo {author} {\bibfnamefont {Z.}~\bibnamefont
			{Li}}, \bibinfo {author} {\bibfnamefont {F.}~\bibnamefont {Tabataba-Vakili}},
		\bibinfo {author} {\bibfnamefont {S.}~\bibnamefont {Zhao}}, \bibinfo {author}
		{\bibfnamefont {A.}~\bibnamefont {Rupp}}, \bibinfo {author} {\bibfnamefont
			{I.}~\bibnamefont {Bilgin}}, \bibinfo {author} {\bibfnamefont
			{Z.}~\bibnamefont {Herdegen}}, \bibinfo {author} {\bibfnamefont
			{B.}~\bibnamefont {M{\"a}rz}}, \bibinfo {author} {\bibfnamefont
			{K.}~\bibnamefont {Watanabe}}, \bibinfo {author} {\bibfnamefont
			{T.}~\bibnamefont {Taniguchi}}, \bibinfo {author} {\bibfnamefont {G.~R.}\
			\bibnamefont {Schleder}}, \bibinfo {author} {\bibfnamefont {A.~S.}\
			\bibnamefont {Baimuratov}}, \bibinfo {author} {\bibfnamefont
			{E.}~\bibnamefont {Kaxiras}}, \bibinfo {author} {\bibfnamefont
			{K.}~\bibnamefont {M{\"u}ller-Caspary}},\ and\ \bibinfo {author}
		{\bibfnamefont {A.}~\bibnamefont {H{\"o}gele}},\ }\bibfield  {title}
	{\bibinfo {title} {Lattice reconstruction in {MoSe2--WSe2} heterobilayers
			synthesized by chemical vapor deposition},\ }\href
	{https://doi.org/10.1021/acs.nanolett.2c05094} {\bibfield  {journal}
		{\bibinfo  {journal} {Nano Letters}\ }\textbf {\bibinfo {volume} {23}},\
		\bibinfo {pages} {4160} (\bibinfo {year} {2023})}\BibitemShut {NoStop}%
	\bibitem [{\citenamefont {Sung}\ \emph {et~al.}(2020)\citenamefont {Sung},
		\citenamefont {Zhou}, \citenamefont {Scuri}, \citenamefont {Z{\'o}lyomi},
		\citenamefont {Andersen}, \citenamefont {Yoo}, \citenamefont {Wild},
		\citenamefont {Joe}, \citenamefont {Gelly}, \citenamefont {Heo} \emph
		{et~al.}}]{sung2020broken}%
	\BibitemOpen
	\bibfield  {author} {\bibinfo {author} {\bibfnamefont {J.}~\bibnamefont
			{Sung}}, \bibinfo {author} {\bibfnamefont {Y.}~\bibnamefont {Zhou}}, \bibinfo
		{author} {\bibfnamefont {G.}~\bibnamefont {Scuri}}, \bibinfo {author}
		{\bibfnamefont {V.}~\bibnamefont {Z{\'o}lyomi}}, \bibinfo {author}
		{\bibfnamefont {T.~I.}\ \bibnamefont {Andersen}}, \bibinfo {author}
		{\bibfnamefont {H.}~\bibnamefont {Yoo}}, \bibinfo {author} {\bibfnamefont
			{D.~S.}\ \bibnamefont {Wild}}, \bibinfo {author} {\bibfnamefont {A.~Y.}\
			\bibnamefont {Joe}}, \bibinfo {author} {\bibfnamefont {R.~J.}\ \bibnamefont
			{Gelly}}, \bibinfo {author} {\bibfnamefont {H.}~\bibnamefont {Heo}}, \emph
		{et~al.},\ }\bibfield  {title} {\bibinfo {title} {Broken mirror symmetry in
			excitonic response of reconstructed domains in twisted mose 2/mose 2
			bilayers},\ }\href@noop {} {\bibfield  {journal} {\bibinfo  {journal} {Nature
				Nanotechnology}\ }\textbf {\bibinfo {volume} {15}},\ \bibinfo {pages} {750}
		(\bibinfo {year} {2020})}\BibitemShut {NoStop}%
	\bibitem [{\citenamefont {Arnold}\ \emph {et~al.}(2023)\citenamefont {Arnold},
		\citenamefont {Ghasemifard}, \citenamefont {Kuc}, \citenamefont {Kunstmann},\
		and\ \citenamefont {Heine}}]{arnold2023relaxation}%
	\BibitemOpen
	\bibfield  {author} {\bibinfo {author} {\bibfnamefont {F.~M.}\ \bibnamefont
			{Arnold}}, \bibinfo {author} {\bibfnamefont {A.}~\bibnamefont {Ghasemifard}},
		\bibinfo {author} {\bibfnamefont {A.}~\bibnamefont {Kuc}}, \bibinfo {author}
		{\bibfnamefont {J.}~\bibnamefont {Kunstmann}},\ and\ \bibinfo {author}
		{\bibfnamefont {T.}~\bibnamefont {Heine}},\ }\bibfield  {title} {\bibinfo
		{title} {Relaxation effects in twisted bilayer molybdenum disulfide:
			structure, stability, and electronic properties},\ }\href@noop {} {\bibfield
		{journal} {\bibinfo  {journal} {2D Materials}\ }\textbf {\bibinfo {volume}
			{10}},\ \bibinfo {pages} {045010} (\bibinfo {year} {2023})}\BibitemShut
	{NoStop}%
	\bibitem [{\citenamefont {Andersen}\ \emph {et~al.}(2021)\citenamefont
		{Andersen}, \citenamefont {Scuri}, \citenamefont {Sushko}, \citenamefont
		{De~Greve}, \citenamefont {Sung}, \citenamefont {Zhou}, \citenamefont {Wild},
		\citenamefont {Gelly}, \citenamefont {Heo}, \citenamefont {B{\'e}rub{\'e}}
		\emph {et~al.}}]{andersen2021excitons}%
	\BibitemOpen
	\bibfield  {author} {\bibinfo {author} {\bibfnamefont {T.~I.}\ \bibnamefont
			{Andersen}}, \bibinfo {author} {\bibfnamefont {G.}~\bibnamefont {Scuri}},
		\bibinfo {author} {\bibfnamefont {A.}~\bibnamefont {Sushko}}, \bibinfo
		{author} {\bibfnamefont {K.}~\bibnamefont {De~Greve}}, \bibinfo {author}
		{\bibfnamefont {J.}~\bibnamefont {Sung}}, \bibinfo {author} {\bibfnamefont
			{Y.}~\bibnamefont {Zhou}}, \bibinfo {author} {\bibfnamefont {D.~S.}\
			\bibnamefont {Wild}}, \bibinfo {author} {\bibfnamefont {R.~J.}\ \bibnamefont
			{Gelly}}, \bibinfo {author} {\bibfnamefont {H.}~\bibnamefont {Heo}}, \bibinfo
		{author} {\bibfnamefont {D.}~\bibnamefont {B{\'e}rub{\'e}}}, \emph {et~al.},\
	}\bibfield  {title} {\bibinfo {title} {Excitons in a reconstructed moir{\'e}
			potential in twisted {WSe2/WSe2} homobilayers},\ }\href@noop {} {\bibfield
		{journal} {\bibinfo  {journal} {Nature Materials}\ }\textbf {\bibinfo
			{volume} {20}},\ \bibinfo {pages} {480} (\bibinfo {year} {2021})}\BibitemShut
	{NoStop}%
	\bibitem [{\citenamefont {Enaldiev}\ \emph {et~al.}(2020)\citenamefont
		{Enaldiev}, \citenamefont {Z{\'o}lyomi}, \citenamefont {Yelgel},
		\citenamefont {Magorrian},\ and\ \citenamefont
		{Fal’ko}}]{enaldiev2020stacking}%
	\BibitemOpen
	\bibfield  {author} {\bibinfo {author} {\bibfnamefont {V.}~\bibnamefont
			{Enaldiev}}, \bibinfo {author} {\bibfnamefont {V.}~\bibnamefont
			{Z{\'o}lyomi}}, \bibinfo {author} {\bibfnamefont {C.}~\bibnamefont {Yelgel}},
		\bibinfo {author} {\bibfnamefont {S.}~\bibnamefont {Magorrian}},\ and\
		\bibinfo {author} {\bibfnamefont {V.}~\bibnamefont {Fal’ko}},\ }\bibfield
	{title} {\bibinfo {title} {Stacking domains and dislocation networks in
			marginally twisted bilayers of transition metal dichalcogenides},\
	}\href@noop {} {\bibfield  {journal} {\bibinfo  {journal} {Physical Review
				Letters}\ }\textbf {\bibinfo {volume} {124}},\ \bibinfo {pages} {206101}
		(\bibinfo {year} {2020})}\BibitemShut {NoStop}%
	\bibitem [{\citenamefont {Carr}\ \emph {et~al.}(2018)\citenamefont {Carr},
		\citenamefont {Massatt}, \citenamefont {Torrisi}, \citenamefont {Cazeaux},
		\citenamefont {Luskin},\ and\ \citenamefont {Kaxiras}}]{carr2018relaxation}%
	\BibitemOpen
	\bibfield  {author} {\bibinfo {author} {\bibfnamefont {S.}~\bibnamefont
			{Carr}}, \bibinfo {author} {\bibfnamefont {D.}~\bibnamefont {Massatt}},
		\bibinfo {author} {\bibfnamefont {S.~B.}\ \bibnamefont {Torrisi}}, \bibinfo
		{author} {\bibfnamefont {P.}~\bibnamefont {Cazeaux}}, \bibinfo {author}
		{\bibfnamefont {M.}~\bibnamefont {Luskin}},\ and\ \bibinfo {author}
		{\bibfnamefont {E.}~\bibnamefont {Kaxiras}},\ }\bibfield  {title} {\bibinfo
		{title} {Relaxation and domain formation in incommensurate two-dimensional
			heterostructures},\ }\href@noop {} {\bibfield  {journal} {\bibinfo  {journal}
			{Physical Review B}\ }\textbf {\bibinfo {volume} {98}},\ \bibinfo {pages}
		{224102} (\bibinfo {year} {2018})}\BibitemShut {NoStop}%
	\bibitem [{\citenamefont {Ferreira}\ \emph {et~al.}(2021)\citenamefont
		{Ferreira}, \citenamefont {Magorrian}, \citenamefont {Enaldiev},
		\citenamefont {Ruiz-Tijerina},\ and\ \citenamefont
		{Fal'ko}}]{ferreira2021band}%
	\BibitemOpen
	\bibfield  {author} {\bibinfo {author} {\bibfnamefont {F.}~\bibnamefont
			{Ferreira}}, \bibinfo {author} {\bibfnamefont {S.}~\bibnamefont {Magorrian}},
		\bibinfo {author} {\bibfnamefont {V.}~\bibnamefont {Enaldiev}}, \bibinfo
		{author} {\bibfnamefont {D.}~\bibnamefont {Ruiz-Tijerina}},\ and\ \bibinfo
		{author} {\bibfnamefont {V.}~\bibnamefont {Fal'ko}},\ }\bibfield  {title}
	{\bibinfo {title} {Band energy landscapes in twisted homobilayers of
			transition metal dichalcogenides},\ }\href@noop {} {\bibfield  {journal}
		{\bibinfo  {journal} {Applied Physics Letters}\ }\textbf {\bibinfo {volume}
			{118}},\ \bibinfo {pages} {241602} (\bibinfo {year} {2021})}\BibitemShut
	{NoStop}%
	\bibitem [{\citenamefont {Naik}\ \emph {et~al.}(2022)\citenamefont {Naik},
		\citenamefont {Regan}, \citenamefont {Zhang}, \citenamefont {Chan},
		\citenamefont {Li}, \citenamefont {Wang}, \citenamefont {Yoon}, \citenamefont
		{Ong}, \citenamefont {Zhao}, \citenamefont {Zhao} \emph
		{et~al.}}]{naik2022intralayer}%
	\BibitemOpen
	\bibfield  {author} {\bibinfo {author} {\bibfnamefont {M.~H.}\ \bibnamefont
			{Naik}}, \bibinfo {author} {\bibfnamefont {E.~C.}\ \bibnamefont {Regan}},
		\bibinfo {author} {\bibfnamefont {Z.}~\bibnamefont {Zhang}}, \bibinfo
		{author} {\bibfnamefont {Y.-H.}\ \bibnamefont {Chan}}, \bibinfo {author}
		{\bibfnamefont {Z.}~\bibnamefont {Li}}, \bibinfo {author} {\bibfnamefont
			{D.}~\bibnamefont {Wang}}, \bibinfo {author} {\bibfnamefont {Y.}~\bibnamefont
			{Yoon}}, \bibinfo {author} {\bibfnamefont {C.~S.}\ \bibnamefont {Ong}},
		\bibinfo {author} {\bibfnamefont {W.}~\bibnamefont {Zhao}}, \bibinfo {author}
		{\bibfnamefont {S.}~\bibnamefont {Zhao}}, \emph {et~al.},\ }\bibfield
	{title} {\bibinfo {title} {Intralayer charge-transfer moir{\'e} excitons in
			van der waals superlattices},\ }\href@noop {} {\bibfield  {journal} {\bibinfo
			{journal} {Nature}\ }\textbf {\bibinfo {volume} {609}},\ \bibinfo {pages}
		{52} (\bibinfo {year} {2022})}\BibitemShut {NoStop}%
	\bibitem [{\citenamefont {Li}\ \emph {et~al.}(2024)\citenamefont {Li},
		\citenamefont {Xiang}, \citenamefont {Naik}, \citenamefont {Kim},
		\citenamefont {Li}, \citenamefont {Sailus}, \citenamefont {Banerjee},
		\citenamefont {Taniguchi}, \citenamefont {Watanabe}, \citenamefont {Tongay}
		\emph {et~al.}}]{li2024imaging}%
	\BibitemOpen
	\bibfield  {author} {\bibinfo {author} {\bibfnamefont {H.}~\bibnamefont
			{Li}}, \bibinfo {author} {\bibfnamefont {Z.}~\bibnamefont {Xiang}}, \bibinfo
		{author} {\bibfnamefont {M.~H.}\ \bibnamefont {Naik}}, \bibinfo {author}
		{\bibfnamefont {W.}~\bibnamefont {Kim}}, \bibinfo {author} {\bibfnamefont
			{Z.}~\bibnamefont {Li}}, \bibinfo {author} {\bibfnamefont {R.}~\bibnamefont
			{Sailus}}, \bibinfo {author} {\bibfnamefont {R.}~\bibnamefont {Banerjee}},
		\bibinfo {author} {\bibfnamefont {T.}~\bibnamefont {Taniguchi}}, \bibinfo
		{author} {\bibfnamefont {K.}~\bibnamefont {Watanabe}}, \bibinfo {author}
		{\bibfnamefont {S.}~\bibnamefont {Tongay}}, \emph {et~al.},\ }\bibfield
	{title} {\bibinfo {title} {Imaging moir{\'e} excited states with photocurrent
			tunnelling microscopy},\ }\href@noop {} {\bibfield  {journal} {\bibinfo
			{journal} {Nature materials}\ }\textbf {\bibinfo {volume} {23}},\ \bibinfo
		{pages} {633} (\bibinfo {year} {2024})}\BibitemShut {NoStop}%
	\bibitem [{\citenamefont {Enaldiev}\ \emph {et~al.}(2021)\citenamefont
		{Enaldiev}, \citenamefont {Ferreira}, \citenamefont {Magorrian},\ and\
		\citenamefont {Fal’ko}}]{Enaldiev_2021}%
	\BibitemOpen
	\bibfield  {author} {\bibinfo {author} {\bibfnamefont {V.~V.}\ \bibnamefont
			{Enaldiev}}, \bibinfo {author} {\bibfnamefont {F.}~\bibnamefont {Ferreira}},
		\bibinfo {author} {\bibfnamefont {S.~J.}\ \bibnamefont {Magorrian}},\ and\
		\bibinfo {author} {\bibfnamefont {V.~I.}\ \bibnamefont {Fal’ko}},\
	}\bibfield  {title} {\bibinfo {title} {Piezoelectric networks and
			ferroelectric domains in twistronic superlattices in {WS2/MoS2} and
			{WSe2/MoSe2} bilayers},\ }\href {https://doi.org/10.1088/2053-1583/abdd92}
	{\bibfield  {journal} {\bibinfo  {journal} {2D Materials}\ }\textbf {\bibinfo
			{volume} {8}},\ \bibinfo {pages} {025030} (\bibinfo {year}
		{2021})}\BibitemShut {NoStop}%
	\bibitem [{\citenamefont {Hagel}\ \emph {et~al.}(2021)\citenamefont {Hagel},
		\citenamefont {Brem}, \citenamefont {Linder\"alv}, \citenamefont {Erhart},\
		and\ \citenamefont {Malic}}]{PhysRevResearch.3.043217}%
	\BibitemOpen
	\bibfield  {author} {\bibinfo {author} {\bibfnamefont {J.}~\bibnamefont
			{Hagel}}, \bibinfo {author} {\bibfnamefont {S.}~\bibnamefont {Brem}},
		\bibinfo {author} {\bibfnamefont {C.}~\bibnamefont {Linder\"alv}}, \bibinfo
		{author} {\bibfnamefont {P.}~\bibnamefont {Erhart}},\ and\ \bibinfo {author}
		{\bibfnamefont {E.}~\bibnamefont {Malic}},\ }\bibfield  {title} {\bibinfo
		{title} {Exciton landscape in van der waals heterostructures},\ }\href
	{https://doi.org/10.1103/PhysRevResearch.3.043217} {\bibfield  {journal}
		{\bibinfo  {journal} {Phys. Rev. Research}\ }\textbf {\bibinfo {volume}
			{3}},\ \bibinfo {pages} {043217} (\bibinfo {year} {2021})}\BibitemShut
	{NoStop}%
	\bibitem [{\citenamefont {Wang}\ \emph {et~al.}(2017)\citenamefont {Wang},
		\citenamefont {Wang}, \citenamefont {Yao}, \citenamefont {Liu},\ and\
		\citenamefont {Yu}}]{wang2017interlayer}%
	\BibitemOpen
	\bibfield  {author} {\bibinfo {author} {\bibfnamefont {Y.}~\bibnamefont
			{Wang}}, \bibinfo {author} {\bibfnamefont {Z.}~\bibnamefont {Wang}}, \bibinfo
		{author} {\bibfnamefont {W.}~\bibnamefont {Yao}}, \bibinfo {author}
		{\bibfnamefont {G.-B.}\ \bibnamefont {Liu}},\ and\ \bibinfo {author}
		{\bibfnamefont {H.}~\bibnamefont {Yu}},\ }\bibfield  {title} {\bibinfo
		{title} {Interlayer coupling in commensurate and incommensurate bilayer
			structures of transition-metal dichalcogenides},\ }\href@noop {} {\bibfield
		{journal} {\bibinfo  {journal} {Physical Review B}\ }\textbf {\bibinfo
			{volume} {95}},\ \bibinfo {pages} {115429} (\bibinfo {year}
		{2017})}\BibitemShut {NoStop}%
	\bibitem [{\citenamefont {Cappelluti}\ \emph {et~al.}(2013)\citenamefont
		{Cappelluti}, \citenamefont {Rold{\'a}n}, \citenamefont {Silva-Guill{\'e}n},
		\citenamefont {Ordej{\'o}n},\ and\ \citenamefont
		{Guinea}}]{cappelluti2013tight}%
	\BibitemOpen
	\bibfield  {author} {\bibinfo {author} {\bibfnamefont {E.}~\bibnamefont
			{Cappelluti}}, \bibinfo {author} {\bibfnamefont {R.}~\bibnamefont
			{Rold{\'a}n}}, \bibinfo {author} {\bibfnamefont {J.}~\bibnamefont
			{Silva-Guill{\'e}n}}, \bibinfo {author} {\bibfnamefont {P.}~\bibnamefont
			{Ordej{\'o}n}},\ and\ \bibinfo {author} {\bibfnamefont {F.}~\bibnamefont
			{Guinea}},\ }\bibfield  {title} {\bibinfo {title} {Tight-binding model and
			direct-gap/indirect-gap transition in single-layer and multilayer
			{MoS$_2$}},\ }\href@noop {} {\bibfield  {journal} {\bibinfo  {journal}
			{Physical Review B}\ }\textbf {\bibinfo {volume} {88}},\ \bibinfo {pages}
		{075409} (\bibinfo {year} {2013})}\BibitemShut {NoStop}%
	\bibitem [{sup()}]{supp}%
	\BibitemOpen
	\href@noop {} {}\bibinfo {note} {See Supplemental Material for further
		results and detials concerning the theoretical model. Supplemental Material
		includes Refs.[43-44]}\BibitemShut {NoStop}%
	\bibitem [{\citenamefont {Khatibi}\ \emph {et~al.}(2018)\citenamefont
		{Khatibi}, \citenamefont {Feierabend}, \citenamefont {Selig}, \citenamefont
		{Brem}, \citenamefont {Linder{\"a}lv}, \citenamefont {Erhart},\ and\
		\citenamefont {Malic}}]{khatibi2018impact}%
	\BibitemOpen
	\bibfield  {author} {\bibinfo {author} {\bibfnamefont {Z.}~\bibnamefont
			{Khatibi}}, \bibinfo {author} {\bibfnamefont {M.}~\bibnamefont {Feierabend}},
		\bibinfo {author} {\bibfnamefont {M.}~\bibnamefont {Selig}}, \bibinfo
		{author} {\bibfnamefont {S.}~\bibnamefont {Brem}}, \bibinfo {author}
		{\bibfnamefont {C.}~\bibnamefont {Linder{\"a}lv}}, \bibinfo {author}
		{\bibfnamefont {P.}~\bibnamefont {Erhart}},\ and\ \bibinfo {author}
		{\bibfnamefont {E.}~\bibnamefont {Malic}},\ }\bibfield  {title} {\bibinfo
		{title} {Impact of strain on the excitonic linewidth in transition metal
			dichalcogenides},\ }\href@noop {} {\bibfield  {journal} {\bibinfo  {journal}
			{2D Materials}\ }\textbf {\bibinfo {volume} {6}},\ \bibinfo {pages} {015015}
		(\bibinfo {year} {2018})}\BibitemShut {NoStop}%
	\bibitem [{\citenamefont {Rostami}\ \emph {et~al.}(2018)\citenamefont
		{Rostami}, \citenamefont {Guinea}, \citenamefont {Polini},\ and\
		\citenamefont {Rold{\'a}n}}]{rostami2018piezoelectricity}%
	\BibitemOpen
	\bibfield  {author} {\bibinfo {author} {\bibfnamefont {H.}~\bibnamefont
			{Rostami}}, \bibinfo {author} {\bibfnamefont {F.}~\bibnamefont {Guinea}},
		\bibinfo {author} {\bibfnamefont {M.}~\bibnamefont {Polini}},\ and\ \bibinfo
		{author} {\bibfnamefont {R.}~\bibnamefont {Rold{\'a}n}},\ }\bibfield  {title}
	{\bibinfo {title} {Piezoelectricity and valley chern number in inhomogeneous
			hexagonal 2d crystals},\ }\href@noop {} {\bibfield  {journal} {\bibinfo
			{journal} {npj 2D Materials and Applications}\ }\textbf {\bibinfo {volume}
			{2}},\ \bibinfo {pages} {15} (\bibinfo {year} {2018})}\BibitemShut {NoStop}%
	\bibitem [{\citenamefont {Enaldiev}\ \emph {et~al.}(2022)\citenamefont
		{Enaldiev}, \citenamefont {Ferreira}, \citenamefont {McHugh},\ and\
		\citenamefont {Fal’ko}}]{enaldiev2022self}%
	\BibitemOpen
	\bibfield  {author} {\bibinfo {author} {\bibfnamefont {V.}~\bibnamefont
			{Enaldiev}}, \bibinfo {author} {\bibfnamefont {F.}~\bibnamefont {Ferreira}},
		\bibinfo {author} {\bibfnamefont {J.}~\bibnamefont {McHugh}},\ and\ \bibinfo
		{author} {\bibfnamefont {V.~I.}\ \bibnamefont {Fal’ko}},\ }\bibfield
	{title} {\bibinfo {title} {Self-organized quantum dots in marginally twisted
			{MoSe2/WSe2} and {MoS2/WS2} bilayers},\ }\href@noop {} {\bibfield  {journal}
		{\bibinfo  {journal} {npj 2D Materials and Applications}\ }\textbf {\bibinfo
			{volume} {6}},\ \bibinfo {pages} {74} (\bibinfo {year} {2022})}\BibitemShut
	{NoStop}%
	\bibitem [{\citenamefont {Sanderson}\ and\ \citenamefont
		{Curtin}(2016)}]{sanderson2016armadillo}%
	\BibitemOpen
	\bibfield  {author} {\bibinfo {author} {\bibfnamefont {C.}~\bibnamefont
			{Sanderson}}\ and\ \bibinfo {author} {\bibfnamefont {R.}~\bibnamefont
			{Curtin}},\ }\bibfield  {title} {\bibinfo {title} {Armadillo: a
			template-based c++ library for linear algebra},\ }\href@noop {} {\bibfield
		{journal} {\bibinfo  {journal} {Journal of Open Source Software}\ }\textbf
		{\bibinfo {volume} {1}},\ \bibinfo {pages} {26} (\bibinfo {year}
		{2016})}\BibitemShut {NoStop}%
	\bibitem [{\citenamefont {Sanderson}\ and\ \citenamefont
		{Curtin}(2019)}]{sanderson2019practical}%
	\BibitemOpen
	\bibfield  {author} {\bibinfo {author} {\bibfnamefont {C.}~\bibnamefont
			{Sanderson}}\ and\ \bibinfo {author} {\bibfnamefont {R.}~\bibnamefont
			{Curtin}},\ }\bibfield  {title} {\bibinfo {title} {Practical sparse matrices
			in c++ with hybrid storage and template-based expression optimisation},\
	}\href@noop {} {\bibfield  {journal} {\bibinfo  {journal} {Mathematical and
				Computational Applications}\ }\textbf {\bibinfo {volume} {24}},\ \bibinfo
		{pages} {70} (\bibinfo {year} {2019})}\BibitemShut {NoStop}%
	\bibitem [{\citenamefont {Weng}\ and\ \citenamefont
		{Gao}(2018)}]{weng2018honeycomb}%
	\BibitemOpen
	\bibfield  {author} {\bibinfo {author} {\bibfnamefont {J.}~\bibnamefont
			{Weng}}\ and\ \bibinfo {author} {\bibfnamefont {S.-P.}\ \bibnamefont {Gao}},\
	}\bibfield  {title} {\bibinfo {title} {A honeycomb-like monolayer of hfo 2
			and the calculation of static dielectric constant eliminating the effect of
			vacuum spacing},\ }\href@noop {} {\bibfield  {journal} {\bibinfo  {journal}
			{Physical Chemistry Chemical Physics}\ }\textbf {\bibinfo {volume} {20}},\
		\bibinfo {pages} {26453} (\bibinfo {year} {2018})}\BibitemShut {NoStop}%
	\bibitem [{\citenamefont {Tang}\ \emph {et~al.}(2022)\citenamefont {Tang},
		\citenamefont {Gu}, \citenamefont {Liu}, \citenamefont {Watanabe},
		\citenamefont {Taniguchi}, \citenamefont {Hone}, \citenamefont {Mak},\ and\
		\citenamefont {Shan}}]{tang2022dielectric}%
	\BibitemOpen
	\bibfield  {author} {\bibinfo {author} {\bibfnamefont {Y.}~\bibnamefont
			{Tang}}, \bibinfo {author} {\bibfnamefont {J.}~\bibnamefont {Gu}}, \bibinfo
		{author} {\bibfnamefont {S.}~\bibnamefont {Liu}}, \bibinfo {author}
		{\bibfnamefont {K.}~\bibnamefont {Watanabe}}, \bibinfo {author}
		{\bibfnamefont {T.}~\bibnamefont {Taniguchi}}, \bibinfo {author}
		{\bibfnamefont {J.~C.}\ \bibnamefont {Hone}}, \bibinfo {author}
		{\bibfnamefont {K.~F.}\ \bibnamefont {Mak}},\ and\ \bibinfo {author}
		{\bibfnamefont {J.}~\bibnamefont {Shan}},\ }\bibfield  {title} {\bibinfo
		{title} {Dielectric catastrophe at the wigner-mott transition in a moir{\'e}
			superlattice},\ }\href@noop {} {\bibfield  {journal} {\bibinfo  {journal}
			{Nature communications}\ }\textbf {\bibinfo {volume} {13}},\ \bibinfo {pages}
		{4271} (\bibinfo {year} {2022})}\BibitemShut {NoStop}%
	\bibitem [{\citenamefont {Brem}\ and\ \citenamefont
		{Malic}(2023)}]{brem2023bosonic}%
	\BibitemOpen
	\bibfield  {author} {\bibinfo {author} {\bibfnamefont {S.}~\bibnamefont
			{Brem}}\ and\ \bibinfo {author} {\bibfnamefont {E.}~\bibnamefont {Malic}},\
	}\bibfield  {title} {\bibinfo {title} {Bosonic delocalization of dipolar
			moir{\'e} excitons},\ }\href@noop {} {\bibfield  {journal} {\bibinfo
			{journal} {Nano Letters}\ }\textbf {\bibinfo {volume} {23}},\ \bibinfo
		{pages} {4627} (\bibinfo {year} {2023})}\BibitemShut {NoStop}%
	\bibitem [{\citenamefont {Brem}\ and\ \citenamefont
		{Malic}(2024)}]{brem2024optical}%
	\BibitemOpen
	\bibfield  {author} {\bibinfo {author} {\bibfnamefont {S.}~\bibnamefont
			{Brem}}\ and\ \bibinfo {author} {\bibfnamefont {E.}~\bibnamefont {Malic}},\
	}\bibfield  {title} {\bibinfo {title} {Optical signatures of moir{\'e}
			trapped biexcitons},\ }\href@noop {} {\bibfield  {journal} {\bibinfo
			{journal} {2D Materials}\ }\textbf {\bibinfo {volume} {11}},\ \bibinfo
		{pages} {025030} (\bibinfo {year} {2024})}\BibitemShut {NoStop}%
\end{thebibliography}
\end{document}


\preprint{APS/123-QED}

\title{Supplemental information: Polarization and charge-separation of moir\'e excitons in van der Waals heterostructures}

\author{Joakim Hagel}
  \affiliation{%
Department of Physics, Chalmers University of Technology, 412 96 Gothenburg, Sweden\\
}%
\author{Samuel Brem}%
\affiliation{%
 Department of Physics, Philipps University of Marburg, 35037 Marburg, Germany\\
}%
  \author{Ermin Malic}%
  \email{ermin.malic@chalmers.se}
  \affiliation{%
 Department of Physics, Philipps University of Marburg, 35037 Marburg, Germany\\
}%
\affiliation{%
Department of Physics, Chalmers University of Technology, 412 96 Gothenburg, Sweden\\
}%

\maketitle
\section{Atomic reconstruction}\label{recon}
When the twist angle is small the lattice no longer remains rigid, but instead undergoes a process known as atomic reconstruction \cite{van2023rotational,weston2020atomic}. Here, the lattice relaxes in order minimize the local stacking energies. Consequently, large domains form which are separated via thin domain walls, which in R-type stacking takes the form of triangular domains \cite{weston2020atomic}. This deformation of the rigid lattice fundamentally changes the geometry and depth of the moir\'e potential \cite{PhysRevMaterials.8.034001}. It is therefore essential to model the relaxation of the lattice in order to gain microscopic insights into the reconstructed moir\'e potential. The atomic reconstruction can be modeled by setting up an integral for the total stacking energy within a continuum model \cite{enaldiev2020stacking,PhysRevMaterials.8.034001}
\begin{equation}
    \mathcal{E}=\int_{\mathcal{A}_{\text{M}}} d^2\mathbf{r}\Big[\sum_{l}\mathcal{U}^{l}+W_{R/H}(\mathbf{r}_0)\Big],
\end{equation}
where $l$ is the layer index, $\mathcal{U}^{l}$ the elastic energy and $W_{R/H}(\mathbf{r}_0)$ the adhesion energy between the layers. Here, $\mathbf{r}_0=\theta \hat{z}\times \mathbf{r}+\mathbf{u}^{t}(\mathbf{r})-\mathbf{u}^{b}(\mathbf{r})$, where $\mathbf{u}(\mathbf{r})$ is the displacement vector responsible for the atomic reconstruction. The elastic energy $\mathcal{U}^{l}$ is given by
\begin{equation}
    \mathcal{U}^{l}=\frac{\lambda^l}{2}(u^{l}_{i,i})^2+\mu^l u^{l}_{i,j}u^{l}_{j,i},
\end{equation}
where $\lambda$ and $\mu$ are the material-specific Lam\'e parameters \cite{ferreira2021band}. Here, we have assumed only small displacements, so that the displacement vector $\mathbf{u}(\mathbf{r})$ can be related to the linear strain tensor as $\varepsilon_{ij}=\frac{1}{2}(u_{i,j}+u_{j,i})$ with $u_{i,j}=\frac{1}{2}(\partial_iu_j+\partial_ju_i)$ and $i(j)=(x,y)$. Furthermore, the adhesion energy between the layers $W_{R/H}(\mathbf{r}_0)$ is given by
\begin{equation}
\begin{split}
        &W_{R/H}(\mathbf{r}_0)=-\kappa \mathcal{Z}_{R/H}^2(\mathbf{r}_0)\\
        &+\sum_{n=0}^{2}\Big[a_1\text{cos}(\mathbf{G}_n\mathbf{r}_0)+a_2\text{sin}(\mathbf{G}_n\mathbf{r}_0+\gamma_{R/H})\Big],
\end{split}
\end{equation}
where $\mathbf{G}_n$ are Brillouin-zone lattice vectors and $\mathcal{Z}_{R/H}^2(\mathbf{r}_0)$ is the term corresponding to the interlayer distance relaxation and is given by
\begin{equation}
\begin{split}
        &\mathcal{Z}_{R/H}(\mathbf{r}_0)=\frac{1}{2\kappa}\sum_{n=0}^{2}\Big[a_1A\text{cos}(\mathbf{G}_n\mathbf{r}_0)\\
        &+a_2|\mathbf{G}_n|\text{sin}(\mathbf{G}_n\mathbf{r}_0+\gamma_{R/H})\Big].
\end{split}
\end{equation}
Here, $\kappa$, $a_1$, $a_2$ and $A$ are parameters fitted from density functional theory (DFT) and obtained from Ref.\cite{enaldiev2020stacking}. By expanding the displacement vectors as a Fourier series $\mathbf{u}^{l}(\mathbf{r})=\sum_n \mathbf{u}^{l}_ne^{i\mathbf{g}_n\mathbf{r}}$, where $\mathbf{g}_n$ are the reciprocal moir\'e vectors, we go from an integral problem depending on $\mathbf{u}^{l}(\bm{r})$ to an optimization problem depending on the Fourier coefficients $\mathbf{u}^{l}_n$, which can then be numerically calculated \cite{PhysRevMaterials.8.034001}. 

\section{Moir\'e potential}\label{sec:pot}
In the twisted MoSe$_2$-WSe$_2$ heterostructure, we have three components to the moir\'e potential for the low-lying KK interlayer excitons \cite{brem2020tunable}. Since the tunneling is very weak around the K-point in R-type stacked MoSe$_2$-WSe$_2$ heterostructures \cite{PhysRevResearch.3.043217}, we can neglect this component to the moir\'e potential. In a atomically reconstructed lattice, the dominating component is instead the scalar strain \cite{van2023rotational}, directly obtained from the displacement vectors 
\begin{equation}\label{eq:scalarStrain}
    S^{l}_{\lambda}(\mathbf{r})=\sum_{i} u^l_{i,i}(\mathbf{r})g^{l}_{\lambda},
\end{equation}
where $\lambda=(c,v)$ is the compound band index and $g^{l}_{\lambda}$ is the gauge factor determining band edge variation as obtained from DFT \cite{khatibi2018impact}. 

In addition to the scalar strain, we also have a smaller component stemming from atomic rotation called the piezo potential \cite{enaldiev2020stacking}. The polarization induced by the piezo charges can be directly obtained from the displacement vectors \cite{PhysRevMaterials.8.034001,enaldiev2020stacking}
\begin{equation}
    \mathbf{\mathcal{P}}^l(\mathbf{r})=e^l_{11}(u^l_{x,x}-u^l_{y,y},-2u^l_{x,y}).
\end{equation}
Here, $e^l_{11}$ is the material-specific piezo coefficient \cite{rostami2018piezoelectricity}. By using the Gauss law and solving the Poisson equation we can calculate the resulting band edge variation from the piezo potential \cite{PhysRevMaterials.8.034001}.

The last component of the moir\'e potential is the stacking-dependent polarization induced shift (alignment shift) of the band edges. By extracting the shift for each high-symmetry stacking from DFT and smoothly interpolating between them \cite{brem2020tunable,hagel2022electrical} we can map out the total potential. Furthermore, by adding the displacement vectors we can deform the geometry of the potential as expected from the the atomic reconstruction \cite{PhysRevMaterials.8.034001}. The final formula for the alignment shift is then given by
\begin{equation}\label{eq:FitFormula}
\begin{split}
    A^{\lambda}_{l}(\mathbf{r})=\text{Re}\Big[v^{\lambda}_l+(\mathcal{A}^{\lambda}_l\\
    +\mathcal{B}_l^{\lambda}e^{i2\pi/3})\sum_{n=0}^{2}e^{i(\mathbf{g}_n\cdot\mathbf{r}+\mathbf{G}_n\cdot \Delta\mathbf{u}^l(\mathbf{r}))}\Big].
\end{split}
\end{equation}
where $v^{\lambda}_l$, $\mathcal{A}^{\lambda}_l$ and $\mathcal{B}^{\lambda}_l$ are parameters which is obtained as DFT input \cite{brem2020tunable}. Here, $\Delta \mathbf{u}(\mathbf{r})=\mathbf{u}^{t}(\mathbf{r})-\mathbf{u}^{b}(\mathbf{r})$ are the displacement vector of the atomic lattice, giving rise to the deformation of the potential.

Since all moir\'e potential components act as a renormalization on the band structure we can merge them into one total moir\'e potential, and then expend it as a Fourier series 
\begin{equation}\label{eq:Expansion}
    V^{\lambda}_{l}(\mathbf{r})=S^{l}_{\lambda}(\mathbf{r})+P^l(\mathbf{r})+A^{\lambda}_{l}(\mathbf{r})=\sum_{\mathbf{g}}m^{l\lambda}_{\bm{g}}e^{i\mathbf{g}\cdot\mathbf{r}},
\end{equation}
where $S^{l}_{\lambda}(\mathbf{r})$ is the scalar strain potential, $P^l(\mathbf{r})$ the piezo potential, $A^{\lambda}_{l}(\mathbf{r})$ the alignment shift, and $V^{\lambda}_{l}(\mathbf{r})$ is the total moir\'e potential. Furthermore, $m^{l\lambda}_{\bm{g}}$ are the Fourier coefficients of the total moir\'e potential as obtained from 
\begin{equation}\label{eq:CoefInt}
      m^{l\lambda}_{\bm{g}}=\frac{1}{\mathcal{A}_{\text{M}}}\int_{\mathcal{A}_{\text{M}}}d\mathbf{r}e^{-i\mathbf{g}\cdot \mathbf{r}}V^{\lambda}_{l}(\mathbf{r})
\end{equation}
with $\mathcal{A}_{\text{M}}$ as the moir\'e unit cell area.

\section{Generalized eigenvalue equation}\label{sec:Ham}
In order to model the exciton energy landscape with charge separation in an atomically reconstructed lattice we first set up a Hamiltonian in second quantization in momentum space
\begin{equation}\label{eq:1}
\begin{split}
H&=\sum_{\bm{k}\lambda}\varepsilon^\lambda_{\bm{k}}\lambda^{\dagger}_{\bm{k}}\lambda_{\bm{k}}+\sum_{\bm{k}\bm{q}\lambda}M^{\lambda}_{\bm{q}}\lambda^{\dagger}_{\bm{k}+\bm{q}}\lambda_{\bm{k}}\\
&+\sum_{\bm{k}\bm{k}^{\prime}\bm{q}}V^{cv}_{\bm{q}}c^{\dagger}_{\bm{k}+\bm{q}}v^{\dagger}_{\bm{k}^{\prime}-\bm{q}}v_{\bm{k}^{\prime}}c_{\bm{k}},
\end{split}
\end{equation}
where the momentum of the electron/hole is given by $\bm{k}^{(\prime)}$, while $\bm{q}$ denotes the transferred momentum. The electron (hole) dispersion is denoted by $\varepsilon^{c(v)}_{\bm{k}}$ and the Fourier-transformed moir\'e potential by $M^{\lambda}_{\bm{q}}$. Furthermore, the Coulomb matrix element responsible for the formation of excitons is given by $V^{cv}_{\bm{q}}$, where the generalized Keldysh potential has been used \cite{erkensten2022microscopic}. Moreover, $c^{(\dagger)}$ and $v^{(\dagger)}$ are the annihilation (creation) operators for the conduction and the valance band, respectively. Since electrons now remain fixed in the molybdenum layer and holes in the tungsten layer, we have dropped the layer index from the equation.

The general two-particle exciton state reads $\ket{X}=\sum_{\bm{k}\bm{k}^{\prime}}\Psi_{\bm{k}\bm{k}^{\prime}}c^{\dagger}_{\bm{k}}v_{\bm{k}^{\prime}}\ket{0}=X^{\dagger}\ket{0}$, where $\Psi_{\bm{k}\bm{k}^{\prime}}$ is the general two-particle wave function and $X^{(\dagger)}$ the exciton annihilation (creation) operator. By acting the Hamiltonian upon the general exciton state we can derive an eigenvalue problem $H\ket{X}=E\ket{X}$, which then reads
\begin{equation}\label{eq:2}
\begin{split}
&(\varepsilon^c_{\bm{k}}-\varepsilon^v_{\bm{k}^{\prime}})\Psi_{\bm{k}\bm{k}^{\prime}}+\sum_{\bm{q}}\Big(M^c_{\bm{q}}\Psi_{\bm{k}-\bm{q},\bm{k}^{\prime}}-M^{v}_{\bm{q}}\Psi_{\bm{k},\bm{k}^{\prime}+\bm{q}}\Big)\\
&-\sum_{\bm{q}}V^{cv}_{\bm{q}}\Psi_{\bm{k}-\bm{q},\bm{k}^{\prime}-\bm{q}}=E\Psi_{\bm{k}\bm{k}^{\prime}},
\end{split}
\end{equation}
where $E$ is the exciton energy. 

By introducing the the center-of-mass momentum (COM) $\bm{Q}=\bm{k}_c-\bm{k}_v=\bm{k}_e+\bm{k}_h$ we can map the dependence on the momentum from $\bm{k}/\bm{k}^{\prime}$ to $\bm{k}/Q$. Here, we have introduced the notation $\Tilde{\Psi}_{\bm{k}\bm{Q}}=\Psi_{\bm{k},\bm{k}-\bm{Q}}\leftrightarrow \Psi_{\bm{k}\bm{k}^{\prime}}=\Tilde{\Psi}_{\bm{k},\bm{k}-\bm{k}^{\prime}}$, which translates between the pictures. Now applying the zone-folding technique we restrict the summation over $\bm{k}/\bm{Q}$ to the first mBZ (mini-Brillouin zone) and mBZ lattice vectors $\bm{g}(\Tilde{\bm{g}})$ as $\Tilde{\Psi}_{\bm{k}+\bm{g},\bm{Q}+\Tilde{\bm{g}}}=\Phi_{\bm{k}\bm{Q}}(\bm{g},\Tilde{\bm{g}})$. Note that the moir\'e potential is periodic such that $M^{\lambda}_{\bm{q}}=\sum_{\bm{g}}m^{\lambda}_{\bm{g}}\delta_{\bm{q},\bm{g}}$, where $m^{\lambda}_{\bm{g}}$ are the Fourier coefficients obtained from \autoref{eq:CoefInt}. Consequently, we can derive the zone-folded eigenvalue problem as 
\begin{equation}
\begin{split}
\label{eq:3}
&\Delta\varepsilon_{\bm{k}\bm{Q}}^{cv}(\bm{g},\Tilde{\bm{g}})\Phi^{\bm{Q}}_{\bm{k}}(\bm{g},\Tilde{\bm{g}})\\
&+\sum_{\bm{g}^{\prime}\Tilde{\bm{g}}^{\prime}}\Big(m^{c}_{\Tilde{\bm{g}}-\Tilde{\bm{g}}^{\prime}}\delta_{\bm{g}-\bm{g}^{\prime},\Tilde{\bm{g}}-\Tilde{\bm{g}}^{\prime}}-m^{v}_{\Tilde{\bm{g}}-\Tilde{\bm{g}}^{\prime}}\delta_{\bm{g}-\bm{g}^{\prime},\Tilde{\bm{g}}-\Tilde{\bm{g}}^{\prime}}\Big)\Phi^{\bm{Q}}_{\bm{k}}(\bm{g}^{\prime},\Tilde{\bm{g}}^{\prime})\\
&-\sum_{\bm{p}\bm{g}^{\prime}}V^{cv}_{\bm{k}-\bm{p}+\bm{g}-\bm{g}^{\prime}}\Phi^{\bm{Q}}_{\bm{k}}(\bm{g}^{\prime},\Tilde{\bm{g}})=E_{\bm{Q}}\Phi^{\bm{Q}}_{\bm{k}}(\bm{g},\Tilde{\bm{g}}),
\end{split}
\end{equation}
where $\Phi^{\bm{Q}}_{\bm{k}}(\bm{g},\Tilde{\bm{g}})$ is the zone-folded two-particle wave function and $E_{\bm{Q}}$ are the exciton energies as a function of COM momentum. Here, $\Delta\varepsilon_{\bm{k}\bm{Q}}^{cv}(\bm{g},\Tilde{\bm{g}})=\varepsilon^c_{\bm{k}+\bm{g}}-\varepsilon^v_{\bm{k}+\bm{g}-\bm{Q}-\Tilde{\bm{g}}}$ is the difference between the zone-folded conduction/valance band dispersion. By treating the eigenvalue problem as a sparse matrix to be diagonalized we can calculate the exciton energies. This was done using the c++ library Armadillo \cite{sanderson2019practical,sanderson2016armadillo}, where the number of moir\'e shells for $\bm{g}$ was converged at 7 and for $\Tilde{\bm{g}}$ it was chosen as $N^{el}_{shell}\approx \frac{\bm{k}^{el}_{cut}}{|\bm{g}_1|}$ such that the number of shells for the electron momentum zone-folding was dynamic with the change in twist angle. Here, we have introduced $\bm{k}^{el}_{cut}=\bm{k}_{max}+\bm{g}_{max}$, where the energies $E_{\bm{Q}}$ were converged at $\bm{k}_{max}\approx 2.2 \text{nm}^{-1}$.

\section{Charge densities}
After diagonalizing the generalized moir\'e exciton eigenvalue problem in \autoref{eq:3} we want to gain access to to the spatial distribution of electrons and holes. This is done by expanding electron densities in real space $\rho^e(\bm{r})=\braket{X_{\bm{Q}}|\Psi_e^{\dagger}(\bm{r})\Psi_e(\bm{r})|X_{\bm{Q}}}$ with the exciton state 
\begin{equation}
    \ket{X_{\bm{Q}}}=\sum_{\bm{k}\bm{g}\Tilde{\bm{g}}}\Phi^{\bm{Q}}_{\bm{k}}(\bm{g},\Tilde{\bm{g}})c^{\dagger}_{\bm{k}+\bm{g}}v_{\bm{k}+\bm{g}-\bm{Q}-\Tilde{\bm{g}}}\ket{0},
\end{equation}
where $\Psi_{\lambda}^{\dagger}(\bm{r})=\sum_{\bm{p}}e^{i\bm{p}\cdot\bm{r}}\lambda^{\dagger}$. Simplifying the the expression for electron densities we obtain
\begin{equation}\label{eq:electron}
    \rho^e(\bm{r})=\sum_{\Delta}e^{i\Delta\cdot\bm{r}}\sum_{\bm{k}\bm{g}\Tilde{\bm{g}}}\Phi^{\bm{Q}*}_{\bm{k}}(\bm{g}+\Delta,\Tilde{\bm{g}}+\Delta)\Phi^{\bm{Q}}_{\bm{k}}(\bm{g},\Tilde{\bm{g}})
\end{equation}
with $\Delta=\bm{g}-\bm{g}^{\prime}$. In analogy, we can also calculate the hole density $\rho^h(\bm{r})=\braket{0|\Psi_v^{\dagger}(\bm{r})\Psi_v(\bm{r})|0}\\
    -\braket{0|X_{\bm{Q}}\Psi_v^{\dagger}(\bm{r})\Psi_v(\bm{r})X^{\dagger}_{\bm{Q}}|0}$ yielding
\begin{equation}\label{eq:hole}
    \rho^h(\bm{r})=\sum_{\Delta}e^{i\Delta\cdot\bm{r}}\sum_{\bm{k}\bm{g}\Tilde{\bm{g}}}\Phi^{\bm{Q}*}_{\bm{k}}(\bm{g},\Tilde{\bm{g}}+\Delta)\Phi^{\bm{Q}}_{\bm{k}}(\bm{g},\Tilde{\bm{g}}).
\end{equation}
By introducing the relative $\bm{r}=\bm{r}_e-\bm{r}_h$ and center-of-mass coordinates $\bm{R}=\frac{1}{\mu}(m_e\bm{r}_e+m_h\bm{r}_h)$, where $\mu$ is the reduced mass, we can set up an expression for the conditional electron-hole densities
\begin{equation}
\begin{split}
    P_{eh}(\bm{r},\bm{R})=	\langle\bm{Q}|\Psi^{\dagger}_e(\bm{R+\beta\bm{r}})\Psi_e(\bm{R}+\beta\bm{r})\\
   \times\Psi^{\dagger}_h(\bm{R}-\alpha\bm{r})\Psi_h(\bm{R}-\alpha\bm{r})|\bm{Q}	\rangle\\
   =\langle0|X_{\bm{Q}}\Psi^{\dagger}_e(\bm{R+\beta\bm{r}})\Psi_e(\bm{R}+\beta\bm{r})\\
   \times\Psi^{\dagger}_h(\bm{R}-\alpha\bm{r})\Psi_h(\bm{R}-\alpha\bm{r})X^{\dagger}_{\bm{Q}}|	0\rangle,
    \end{split}
\end{equation}
where $\alpha(\beta)=m_{e(h)}/(m_e+m_{h})$ and $ P_{eh}(\bm{r},\bm{R})=|\bar{\Psi}_{\bm{Q}}(\bm{r},\bm{R})|^2$ corresponds to the probability. Following the same approach, we can derive an expression for $P_{eh}(\bm{r},\bm{R})$ reading
\begin{equation}
\begin{split}
 P_{eh}(\bm{r},\bm{R})=	\sum_{\substack{\bm{k}\bm{g}\Tilde{\bm{g}}\\\bm{k}^{\prime}\bm{g}^{\prime}\Tilde{\bm{g}}^{\prime}}}e^{i(\Tilde{\bm{g}}-\Tilde{\bm{g}}^{\prime})\cdot\bm{R})}e^{-i(\Tilde{\bm{g}}-\Tilde{\bm{g}}^{\prime})\cdot\alpha\bm{r}}\times\\
 \Phi^{\bm{Q}*}_{\bm{k}}(\bm{g},\Tilde{\bm{g}}) \Phi^{\bm{Q}*}_{\bm{k}}(\bm{g}^{\prime},\Tilde{\bm{g}}^{\prime}).
\end{split}
\end{equation}

By now introducing the unfolded two-particle wave function $\Tilde{\Phi}_{\bm{Q}}(\bm{k}+\bm{g},\Tilde{\bm{g}})=\Phi^{\bm{Q}}_{\bm{k}}(\bm{g},\Tilde{\bm{g}})$ and taking the trace of the relative electron-hole coordinates, i.e integrating over $\bm{r}$ $P_{eh}(\bm{R})=\frac{1}{A}\int d^2r P_{eh}(\bm{r},\bm{R})$, the following expression for the COM density is obtained
\begin{equation}\label{eq:COM}
 P_{eh}(\bm{R})=	\sum_{\Delta^{\prime}}e^{i\Delta^{\prime}\cdot\bm{R}}\sum_{\bm{k}\Tilde{\bm{g}}}\Tilde{\Phi}^{*}_{\bm{Q}}(\bm{k}+\alpha\Delta,\Tilde{\bm{g}}+\Delta)\Tilde{\Phi}_{\bm{Q}}(\bm{k},\Tilde{\bm{g}})
\end{equation}
with $\Delta^{\prime}=\Tilde{\bm{g}}-\Tilde{\bm{g}}^{\prime}$. By comparing \autoref{eq:electron}, \autoref{eq:hole} and \autoref{eq:COM} we can now see that all densities can be rewritten as a function of the same form factor
\begin{equation}
\begin{split}
&\rho_e(\bm{r})=\sum_{\Delta}e^{i\Delta\cdot\bm{r}}\Gamma_{\bm{Q}}(\Delta,\Delta)\\
&\rho_h(\bm{r})=\sum_{\Delta}e^{i\Delta\cdot\bm{r}}\Gamma_{\bm{Q}}(0,\Delta)\\
&\rho_X(\bm{R})=\sum_{\Delta^{\prime}}e^{i\Delta^{\prime}\cdot\bm{R}}\Gamma_{\bm{Q}}(\alpha \Delta^{\prime},\Delta^{\prime}),\\[6pt]
\end{split}
\end{equation}
where $\rho_X(\bm{R})= P_{eh}(\bm{R})$. The form factor is given by
\begin{equation}
\begin{split}\Gamma_{\bm{Q}}(\bm{q},\Delta)=\sum_{\bm{k}\bm{G}}\Tilde{\Phi}_{\bm{Q}}^{*}(\bm{k}+\bm{q},\bm{G}+\Delta)\Tilde{\Phi}_{\bm{Q}}(\bm{k},\bm{G}).
\end{split}
\end{equation}